\documentclass[graybox]{svmult}

\usepackage{type1cm}        
\usepackage{makeidx}         
\usepackage{graphicx}        
\usepackage{multicol}        
\usepackage[bottom]{footmisc}

\usepackage{newtxtext}       %
\usepackage[varvw]{newtxmath}       

\usepackage[utf8]{inputenc}
\usepackage[T1]{fontenc}
\usepackage[english]{babel}
\usepackage{amsmath,amsthm,amsfonts}
\usepackage{multirow}
\usepackage{tikz}
\usepackage{verbatim}
\usepackage{todonotes}
\usepackage{footmisc}
\usepackage{tocloft} 
\usepackage{isotope}
\usepackage{braket}
\usepackage{siunitx}
\usepackage{enumitem}
\usepackage{xcolor}
\usepackage{dsfont}
\DeclareSIUnit\torr{Torr} 
\DeclareSIUnit\gauss{G}

\usepackage{hyperref}
\hypersetup{
    colorlinks=true,
    linkcolor=blue,
    filecolor=blue,      
    urlcolor=cyan,
    citecolor=blue,
    pdftitle={Contribution Title \today},
    pdfpagemode=FullScreen,
    }

\makeindex             

\newcommand{\micro}[1]{\ensuremath{\upmu\mathrm{#1}}}

\newcommand{\abs}[1]{\left\lvert#1\right\rvert}
\renewcommand{\vec}[1]{\ensuremath{\mathbf{#1}}}
\newcommand{\uvec}[1]{\ensuremath{\mathbf{\hat{#1}}}}

\newcommand{\avg}[1]{\ensuremath{\left\langle{#1}\right\rangle}}

\newcommand{\up}{\ensuremath{\ket{\uparrow}}}
\newcommand{\down}{\ensuremath{\ket{\downarrow}}}

\newcommand{\Hjc}{\ensuremath{H_\mathrm{JC}}}
\newcommand{\EvacVec}{\ensuremath{\boldsymbol{\mathcal{E}}_\mathrm{vac}}}
\newcommand{\Evac}{\ensuremath{\mathcal{E}_\mathrm{vac}}}
\newcommand{\exc}{\ensuremath{e}}
\newcommand{\gnd}{\ensuremath{g}}
\newcommand{\Gcav}{\ensuremath{\Gamma_\mathrm{cav}}}
\newcommand{\Gsc}{\ensuremath{\Gamma_\mathrm{sc}}}

\newcommand{\Odisp}{\ensuremath{\Omega_1}}
\newcommand{\tkappa}{\ensuremath{\tilde{\kappa}}}
\newcommand{\Htc}{\ensuremath{H_\mathrm{TC}}}

\newcommand{\Htwist}{\ensuremath{H_\mathrm{twist}}}

\newcommand{\F}{\ensuremath{\mathcal{F}}}

\newcommand{\QFI}{\ensuremath{\mathcal{I}}}
\newcommand{\Hxxzh}{\ensuremath{H_\mathrm{XXZ,h}}}
\newcommand{\Jprog}{\ensuremath{\mathcal{J}}}
\newcommand{\HSY}{\ensuremath{H_\mathrm{SY}}}

\newcommand{\Heff}{\ensuremath{\tilde{H}_\mathrm{XY}}}

\newcommand{\theabstract}{These lecture notes discuss applications of atom-light interactions in cavities to quantum metrology, simulation, and computation.  A focus is on nonlocally interacting spin systems realized by coupling many atoms to a delocalized mode of light.  We will build up from the fundamentals: understanding how a cavity enables light to coherently imprint information on atoms and atoms to imprint information on the light, enabling quantum non-demolition measurements that constitute a powerful means of engineering nonclassical states. By extension, letting the intracavity light act back on the atoms enables coherent photon-mediated interactions.  I start by discussing collective spin models, emphasizing applications in entanglement-enhanced metrology, before proceeding to richer many-body physics enabled by incorporating spatiotemporal control or employing multiple cavity modes.  I will highlight opportunities for leveraging these tools for quantum simulations inspired by problems in condensed matter and quantum gravity.  Along the way, I provide a pedagogical introduction to criteria for strong atom-light coupling, illustrate how the corresponding figure of merit --- the cooperativity --- sets fundamental limits on the coherence of atom-light interactions, and discuss prospects for harnessing high-cooperativity cavity QED in quantum simulation and computation.}

\begin{document}

\title*{Interfacing Atomic Spins with Photons for Quantum Metrology, Simulation and Computation}
\titlerunning{Interfacing Atomic Spins with Photons for Q. Metrology, Simulation \& Computation}
\author{Monika Schleier-Smith\orcidID{0000-0002-4686-3528}}
\institute{Monika Schleier-Smith \at Department of Physics, Stanford University (Stanford, CA, USA), \\\email{schleier@stanford.edu}}
%
%
\maketitle

\abstract*{\theabstract}

\abstract{\theabstract}

\section{Introduction}\label{sec:introduction}
Central to the entire field of \textit{quantum computing and simulation with atoms} --- the theme of this Enrico Fermi Summer School --- is the ability to precisely manipulate atoms with light.  Often we can consider the light as a classical field --- a strong laser beam used for optically trapping atoms or manipulating their internal states, with so many photons that quantum fluctuations can be neglected.  My lectures will instead explore a regime where we harness the quantum properties of light.  To aid in accessing this quantum regime, it is advantageous to enhance the strength of interaction between light and atoms by trapping the atoms together with photons between two mirrors that form a cavity. In this configuration, even a single photon may interact strongly with a single atom.

Strong atom-photon coupling is a powerful enabling tool for quantum science applications that benefit from the complementary advantages of matter and light for storing and distributing quantum information.  While atomic qubits are well suited to processing information via local gates, photons are easy to delocalize and to transmit across long distances.  In one illustrative application, coupling many atoms to a single delocalized mode of light provides a means of efficiently generating collective entangled states, such as squeezed spin states that are resources for enhanced metrology~\cite{schleier2010states,leroux2010implementation,leroux2010orientation,chen2011conditional,bohnet2014reduced,hosten2016measurement,hosten2016quantum,cox2016deterministic,pedrozo2020entanglement,cooper2024graph,huang2023observing,robinson2024direct}.  This same approach opens new opportunities in quantum simulation: letting photons mediate non-local interactions allows, for example, for mimicking the behavior of phonons in solids~\cite{guo2021optical} or realizing exotic interaction graphs featuring in models of quantum gravity~\cite{bentsen2019treelike,periwal2021programmable}.  Photons can also provide new ways of realizing quantum gates between atomic qubits~\cite{welte2018photon} or themselves act as ``flying qubits'' ideally suited to distributing quantum information in future modular computing architectures~\cite{ramette2022any} or across long-distance networks~\cite{dhordjevic2021entanglement,hartung2024quantum}.

These lectures will start from the fundamentals: what is meant by strong atom-light coupling, and how do we achieve it? I will describe how coherent atom-photon interactions enable quantum non-demolition measurements of light using atoms as probes or, conversely, measurements of atoms using light as a probe.  Such measurements are a powerful tool for engineering a range of non-classical states, from Schr\"{o}dinger cat states of light to squeezed states of atomic ensembles to Bell states of atom pairs.  We will further explore how light can be used to mediate coherent interactions between atoms, opening up new avenues in quantum simulation.  Here, we will start with collective, all-to-all interactions that naturally arise when many atoms are coupled to a single mode of light, then proceed to recent progress in making these interactions increasingly programmable.  Along the way, I will highlight applications in quantum metrology and simulation and touch on opportunities in quantum computation.  While examples from quantum metrology will illustrate the unique strength of cavity-QED platforms for scalable generation of collective entangled states, I will also touch on the challenge of --- and prospects for --- operating deep in the strong coupling regime to enable high-fidelity pairwise entangling gates and to access strongly correlated many-body states.

\section{Fundamentals of Strong Atom-Light Coupling}
\subsection{The Jaynes-Cummings Model, Vacuum Rabi Oscillations, and Strong Coupling}
To introduce the concept of strong atom-light coupling, let us consider a simple model of a two-level atom interacting with a cavity mode.   An atom initially in an excited state $\ket{\exc}$ can transition to the ground state $\ket{\gnd}$ by emitting a photon into the cavity.  For a cavity mode of frequency $\omega_c$ that is near-resonant with the atomic transition of frequency $\omega_0 = \omega_c + \Delta$, the system is described by the Jaynes-Cummings Hamiltonian~\cite{jaynes2005comparison}
\begin{equation}\label{eq:Hjc}
\Hjc = g\left(a^{\dagger}s^{-} + a s^{+}\right) + \Delta s^{z},
\end{equation}
where $\vec{s}$ is the spin-$1/2$ operator representing the two-level atom, and we have set $\hbar=1$.  We represent the cavity as a harmonic oscillator and the photons as excitations of this oscillator, with bosonic annihilation (creation) operators $a^{(\dagger)}$.  The coupling $g$ between the spin and the oscillator arises physically from the dipole interaction between the atom and the vacuum field in the cavity, and is thus given by $g = \vec{d}_{\exc\gnd} \cdot \EvacVec$, where $\vec{d}_{\exc\gnd}$ is the dipole matrix element of the atomic transition and $\EvacVec$ is the vacuum field in the cavity, with magnitude $\Evac = \sqrt{\frac{\hbar \omega_c}{2\epsilon_0 \mathcal{V}}}$ for a cavity of mode volume $\mathcal{V}$~\cite{raimond2006exploring,tanji2011interaction}. 

The atom-light coupling $g$ gives rise to \textit{vacuum Rabi oscillations} in which, for a cavity initialized in the vacuum state, an atomic excitation is coherently converted into a photon and back to an atomic excitation.  More generally, $\Hjc$ couples the state with an atom excited and $n$ photons in the cavity to the state $\ket{\gnd,n+1}$ with a ground-state atom and $n+1$ intracavity photons, leading the system to oscillate between these states at frequency $\Omega_{\mathrm{JC},n} = 2\bra{\gnd,n+1}\Hjc\ket{\exc,n} = 2g\sqrt{n}$.  A seminal experiment observing these vacuum Rabi oscillations was performed in a microwave cavity in the group of Serge Haroche \cite{brune1996quantum}.

\begin{figure}[htb]
\sidecaption
\includegraphics[width=0.6\textwidth]{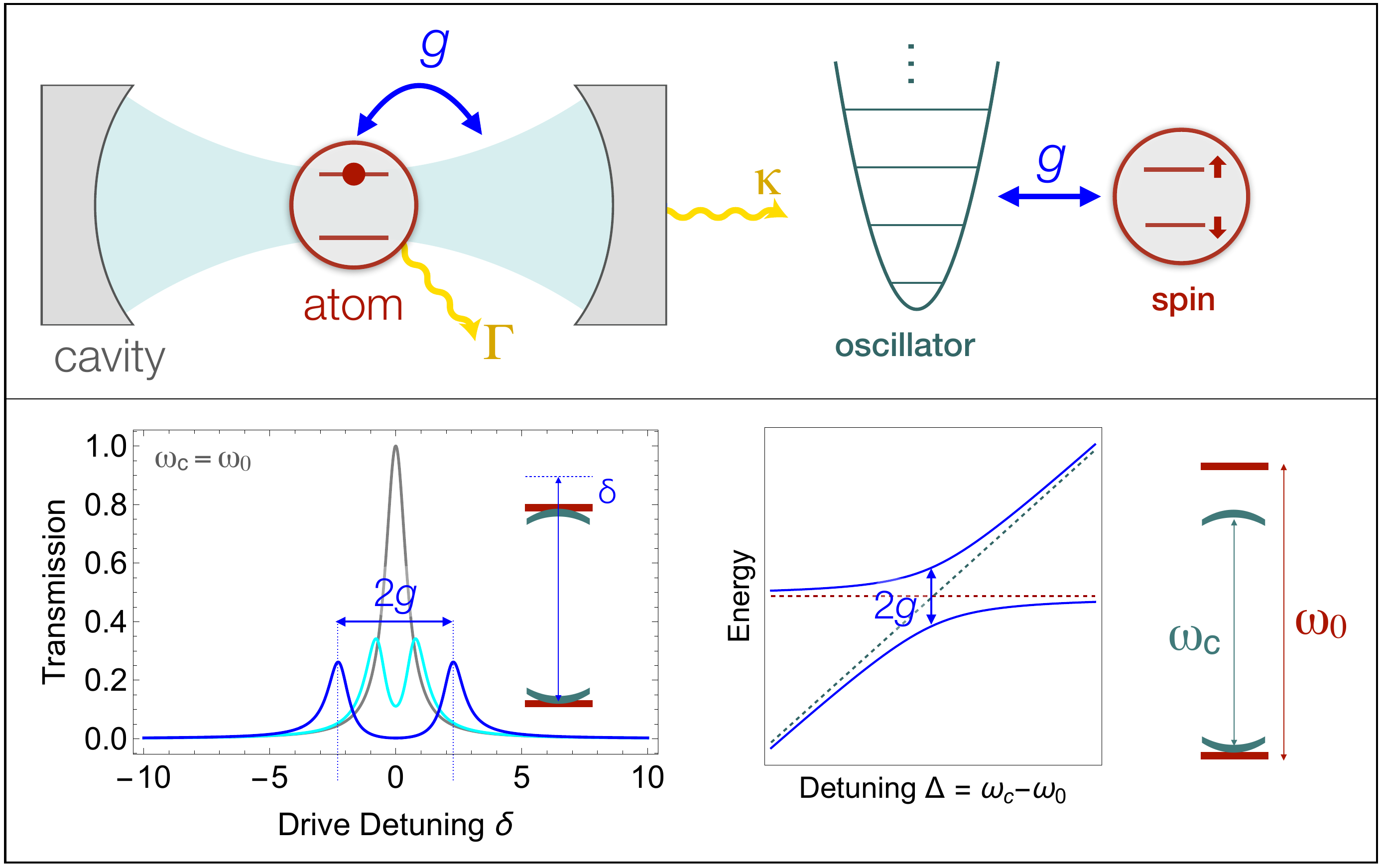}
\caption{A two-level atom (spin) coupled to a cavity (oscillator), as described by the Jaynes-Cummings model of Eq.~\ref{eq:Hjc}.  Bottom left: vacuum Rabi splitting in the transmission spectrum for cavity tuned on resonance with the atomic transition ($\omega_c = \omega_0$).  Bottom right: energy spectrum vs cavity-atom detuning $\Delta$.}
\label{fig:jaynes_cummings}       
\end{figure}

While vacuum Rabi oscillations provide evidence of coherent atom-photon interactions in the time domain, a complementary signature can be found by examining the spectrum of the coupled atom-cavity system in the frequency domain.  In particular, when we examine the energy spectrum as a function of the atom-cavity detuning, we find an avoided crossing of size $2g$, the \textit{vacuum Rabi splitting}.  Observing two well-resolved peaks with splitting $2g$ is often considered the signature of the \textit{strong coupling regime}, as it indicates that the atom-light coupling is stronger than the characteristic decay rates that set the widths of the two peaks.  The vacuum Rabi splitting due to a single atom in an optical cavity was observed in a seminal experiment in the group of J. Kimble~\cite{thompson1992observation}.

\subsection{Figure of Merit: Cooperativity}

The coherence of atom-light interactions depends not only on the absolute strength of coupling $g$ but on the ratio of coupling strength to characteristic decay rates.  Observing the vacuum Rabi splitting requires that the peaks be well separated relative to their widths, set by the arithmetic mean $(\kappa + \Gamma)/2$ of two decay rates: the rate $\kappa$ of photon leakage from the cavity and the spontaneous decay rate $\Gamma$ of an excited atom.  However, a more fundamental criterion for strong coupling is set by comparing the vacuum Rabi frequency to the geometric mean $\sqrt{\kappa\Gamma}$ of the decay rates~\cite{kimble1998strong,tanji2011interaction}, leading to a single dimensionless figure of merit known as the cooperativity $\eta = \frac{4g^2}{\kappa \Gamma}$ (also often denoted $C$).  Specifically, the cooperativity compares the rate at which an excited atom decays by emitting a photon into the cavity, given by Fermi's golden rule as $\Gcav = 4g^2/\kappa$, to the spontaneous emission rate $\Gamma$ of an atom in free space.  Thus, the cooperativity quantifies how well light from a single atom is channeled into a single cavity mode.

The cooperativity can equivalently be expressed purely in terms of geometrical parameters~\cite{tanji2011interaction}.  For a two-level atom or (equivalently) a cycling transition, the emission rates into the cavity ($\Gcav$) and into free space ($\Gamma$) have identical dependences on the transition dipole matrix element, which thus factors out of the ratio $\Gcav/\Gamma$.  We are left with
\begin{equation}
\eta = \frac{\Gcav}{\Gamma} = \frac{4g^2}{\kappa \Gamma} = \frac{24F}{\pi k^2 w_0^2},
\end{equation}
where $F$ is the finesse, $k$ is the wavenumber of the cavity mode, and $w_0$ is the mode waist.  In particular, the expression on the right-hand side is proportional to the ratio of the atomic cross-section $\sigma \propto 1/k^2$ to the cross-sectional area of the cavity mode, enhanced by a factor $F/\pi$ quantifying the average number of round trips made by a photon in the cavity.  

Whereas we so far focused on the \textit{single-atom} cooperativity, certain applications benefit from a collective enhancement from operating with $N$ atoms.  For $N$ atoms, the vacuum Rabi splitting is increased to $g\sqrt{N}$, leading to a collective cooperativity $N\eta$.  Intuitively, the collective enhancement arises because the fields scattered by different atoms into the cavity constructively interfere, leading to an $N$-fold enhancement in the ratio $\Gcav/\Gamma$ of the scattering rate into the cavity compared to that into free space.

\begin{trailer}{Collective Enhancement: Superradiant Scattering}

The superradiant scattering at the heart of collectively enhanced atom-light interactions has recently been observed and controlled at the microscopic level.  Figure~\ref{fig:superradiance} shows an experiment by Yan \textit{et al.}~\cite{yan2023superradiant} in which optical tweezers were used to precisely position individual atoms relative to the standing wave of light in the cavity~\cite{deist2022superresolution,deist2022mid}. For an array of $N$ atoms spaced by an integer number of wavelengths, they observed constructive interference in the scattering into the cavity, resulting in a photon emission rate scaling as (nearly) $N^2$, whereas the scattering from atoms separated by a half-integer spacing destructively interfered.  

\begin{figure}[htb]
\includegraphics[width=\textwidth]{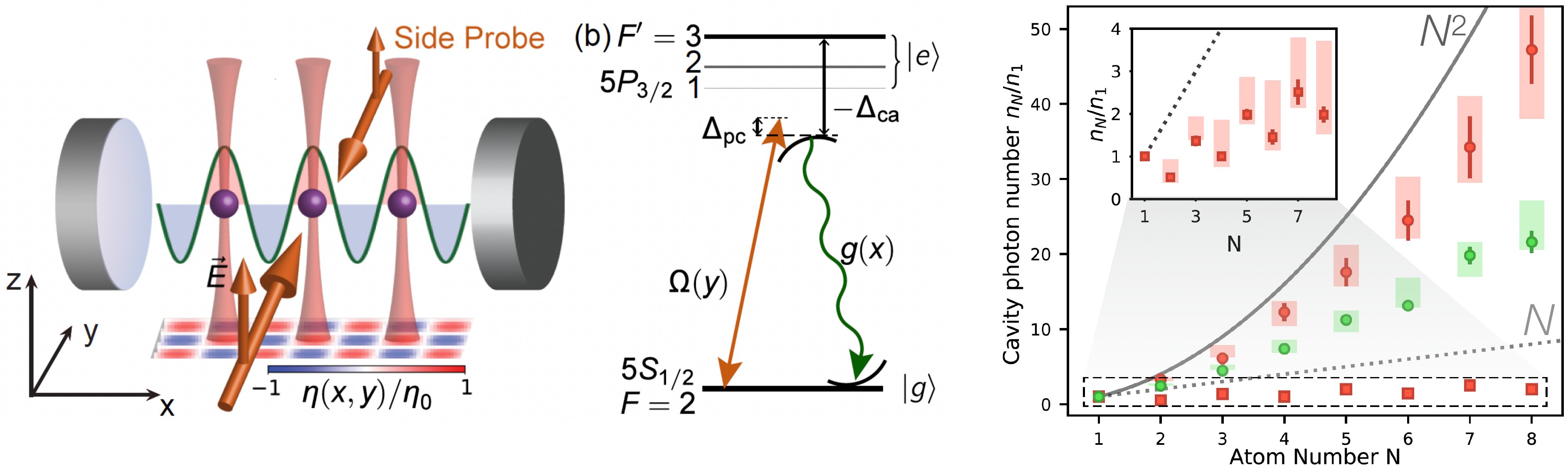}
\caption{\textit{Adapted from Yan et al.~\cite{yan2023superradiant}}.  \textbf{Super- and sub-radiant scattering} observed by positioning atoms in optical tweezers  with subwavelength precision within the standing-wave mode of a Fabry-Perot cavity. Left: schematic of the experimental setup.  Right: dependence of cavity photon number on atom number $N$ for tweezers spaced by integer (red circles) and half-integer (red squares) wavelengths.}\label{fig:superradiance}
\end{figure}
\end{trailer}

\subsection{Cavity Technology}

\begin{figure}[b]
\includegraphics[width=\textwidth]{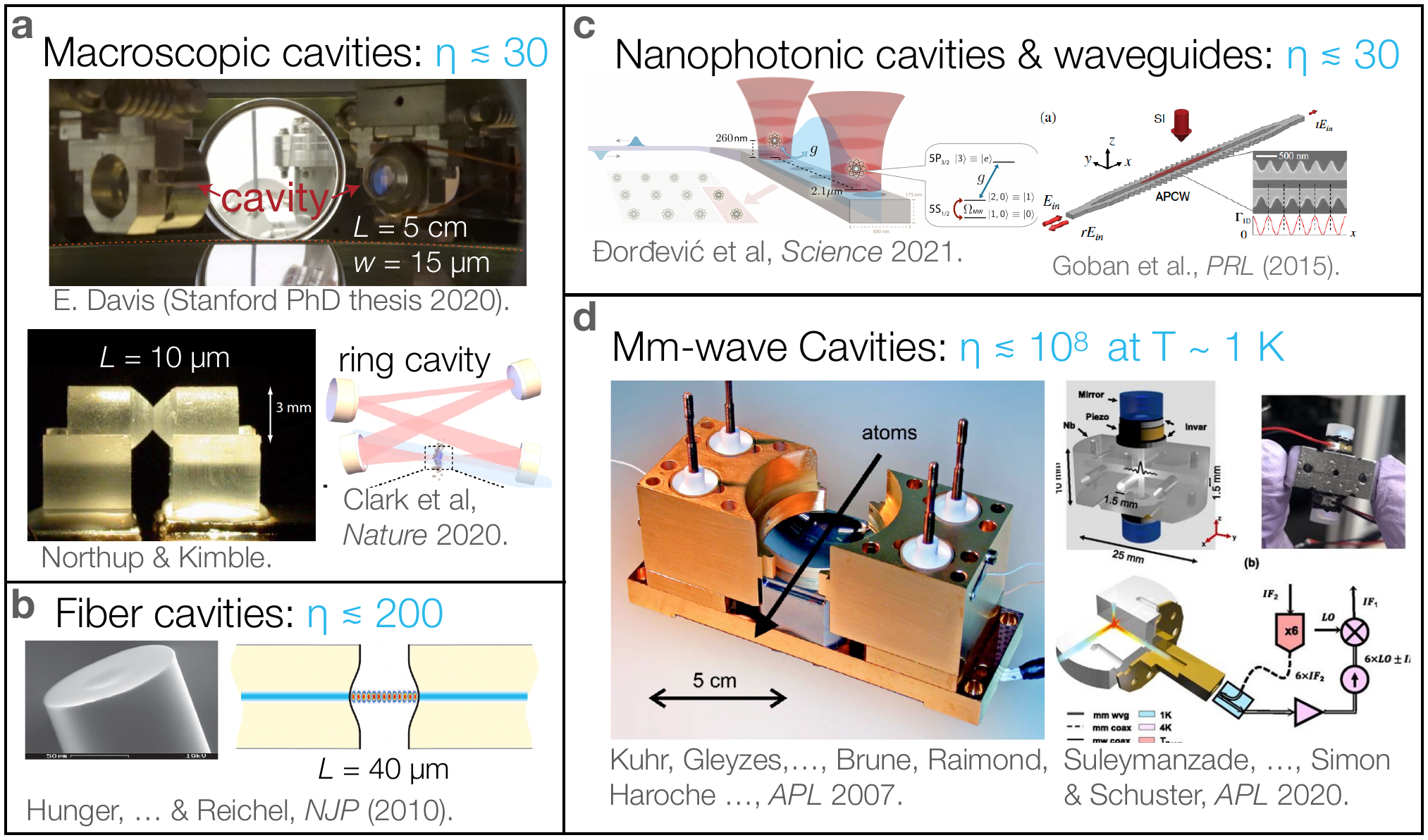}
\caption{\textbf{Cavities come in many forms: examples of cavity types for achieving strong coupling $\eta> 1$.} (a) Optical cavities formed by macroscopic super polished mirrors, including near-concentric~\cite{davis2020engineering} and near-planar~\cite{northup2008coherent} Fabry-Perot cavities and ring cavity~\cite{clark2020observation}.  (b) Optical fiber cavities~\cite{hunger2010fiber}.  (c) Nanophotonic cavities and waveguides~\cite{dhordjevic2021entanglement,goban2015superradiance}.  (c) Millimeter-wave cavities, including Fabry-Perot cavity~\cite{kuhr2007ultrahigh} and three-dimensional cavity~\cite{suleymanzade2020tunable}.}\label{fig:cavity_examples}
\end{figure}

To access the strong-coupling regime, seminal experiments focused on optical cavity designs featuring a small mode volume $\mathcal{V}$, motivated by the scaling $g\propto 1/\sqrt{\mathcal{V}}$.  Illustrative examples include the 10-$\micro{m}$-long cavity from the PhD thesis of T. Northup in the Kimble group~\cite{northup2008coherent}, shown in Fig.~\ref{fig:cavity_examples}; a 40-$\micro{m}$-long optical fiber cavity fabricated by CO$_2$ laser machining in the group of J.~Reichel~\cite{hunger2010fiber}, with cooperativity $\eta\approx 200$; and nanophotonic cavities that confine light to mode volumes on the wavelength scale~\cite{goban2015superradiance, dhordjevic2021entanglement}, achieving cooperativities as high as $\eta\approx 30$~\cite{dhordjevic2021entanglement}.

While small mode volume is necessary for applications that require a well-resolved vacuum Rabi splitting ($g \gg \kappa + \Gamma$), we have argued that a more fundamental criterion for strong coupling is a high cooperativity ($g > \sqrt{\kappa \Gamma}$).  The cooperativity $\eta = 4g^2/(\kappa\Gamma)$ depends only on the cross-sectional area and not on the length $L$, since increasing the length decreases both $g^2 \propto 1/L$ and $\kappa \propto 1/L$. Thus, recent experiments have explored benefits of longer cavities~\cite{stute2012toward,davis2020engineering,bolognini2025new}.  In particular, centimeter-scale cavity lengths are advantageous for experiments with trapped ions or Rydberg atoms to avoid proximity to charges on dielectric mirrors; and, more broadly, they
provide optical access for trapping and addressing arrays of single atoms~\cite{deist2022mid,deist2022superresolution}, ions~\cite{stute2012toward,casabone2015enhanced}, or atomic ensembles~\cite{periwal2021programmable, cooper2024graph}.  Even in such long cavities, operating near the concentric limit permits small mode waists $w_0 \approx 15~\micro{m}$~\cite{stute2012toward,davis2019photon,davis2020engineering,bolognini2025new}, for which a modest finesse $F\sim 10^4$ suffices to access the strong coupling regime $\eta > 1$ at near-infrared wavelengths.

Just as there is no single standardized cavity design, there are also a variety of ways of confining atoms in cavities.  For experiments with atomic ensembles in optical cavities, a common approach is to trap atoms in an intracavity standing wave to ensure their localization within the cavity mode.  Homogeneous atom-cavity coupling is then achieved by choosing the trapping wavelength $\lambda_\mathrm{trap}$ to be commensurate with wavelength $\lambda_0$ of the atomic transition that couples to the cavity --- for example, $\lambda_\mathrm{trap} = 2\lambda_0$ with $\lambda_0 = 780$~nm for the rubidium D2 line~\cite{lee2014many}.  Incorporating transverse trapping beams allows for positioning arrays of atom clouds~\cite{periwal2021programmable,cooper2024graph,kroeze2025directly} or of single atoms~\cite{welte2018photon,deist2022superresolution,hartung2024quantum} within cavities.  For spin physics, it suffices to operate with thermal atoms, provided that the atoms are sufficiently cold and tightly trapped to avoid decoherence from spin-motion coupling.  However, degenerate Bose and Fermi gases have also been trapped in cavities, notably for exploration of self-organized phases~\cite{baumann2010dicke,kroeze2018spinor,zhang2021observation,helson2023density}.

While we have so far focused on introducing optical cavities, significantly higher cooperativity has been achieved for microwave or millimeter-wave photons~\cite{kuhr2007ultrahigh,deleglise2008reconstruction}.  Here, the long wavelength provides relative insensitivity to surface roughness of the mirrors, and superconducting mirrors have allowed for achieving high finesse $F > 10^9$.  Taking advantage of these features, the group of S.~Haroche developed a superconducting Fabry-Perot cavity achieving a cooperativity $\eta = 4\times 10^8$ for circular Rydberg atoms~\cite{kuhr2007ultrahigh,deleglise2008reconstruction}.

The millimeter-wave regime introduces additional experimental complexity, as superconducting cavities require operating at cryogenic temperature~\cite{kuhr2007ultrahigh,suleymanzade2020tunable}.  The pioneering experiments by S. Haroche operated with a beam of velocity-selected atoms traversing a superconducting Fabry-Perot cavity~\cite{deleglise2008reconstruction}.  In ongoing work, we have developed a similar cavity optimized for combining high cooperativity with optical access~\cite{zhang2025optically}, with the goal of trapping arrays of single atoms within the millimeter-wave cavity.  A step in this direction are recent experiments featuring an ensemble of cold atoms trapped in a three-dimensional millimeter-wave cavity crossed by an optical cavity, which enabled high-efficiency quantum transduction~\cite{kumar2023quantum}.

\section{Quantum Non-Demolition (QND) Measurements}\label{sec:qnd_measurement}

One powerful class of techniques enabled by atom-light interactions in optical cavities are \textit{quantum non-demolition (QND) measurements}.  By coupling atoms with light and using one subsystem --- either the light or the atoms --- as a probe, the other subsystem may be projected into a non-classical state.  Applications include engineering Fock states and Schr\"{o}dinger cat states of light~\cite{guerlin2007progressive,deleglise2008reconstruction}, generating squeezed states~\cite{schleier2010states,hosten2016measurement,bohnet2014reduced,cox2016deterministic} and Dicke states~\cite{mcconnell2013generating,haas2014entangled,mcconnell2015entanglement} of collective atomic spins, and generating Bell states of atom pairs~\cite{welte2017cavity,dhordjevic2021entanglement}.

A quantum non-demolition measurement is one that achieves the platonic ideal of leaving the measured system in a state corresponding to the measurement outcome~\cite{braginsky1980quantum,raimond2006exploring}.  In practice, this is a nontrivial feat: for example, once we have detected a photon via the photoelectric effect, the photon is no longer there.  The same would be true even if we used an atom to absorb a photon in the controlled setting of a vacuum Rabi oscillation in a cavity.  However, a QND interaction can be achieved by detuning the cavity from resonance with the atomic transition~\cite{raimond2006exploring}, so that the intracavity light interacts with the atom via the ac Stark effect. The resulting spectrum of the coupled atom-cavity system, obtained in second-order perturbation theory from the Jaynes-Cummings model, is
\begin{equation}\label{eq:Hd}
H_d = \frac{g^2}{\Delta}\,a^\dagger a\, \sigma^z .
\end{equation}
In this \textit{dispersive regime} of cavity QED, the frequency of the cavity mode is shifted by an amount $\pm g^2/\Delta$ depending on the atomic state --- or, equivalently, the atomic spin precesses at a frequency proportional to the intracavity photon number $n = a^\dagger a$.  Importantly, only the phase of the coupled light-matter system evolves under $H_d$, while the number of atomic and photonic excitations is preserved.  Thus, $H_d$ enables QND measurements of photon number $n$ by using the atom as a probe; or of the spin $\sigma^z$ by using the light as a probe.

\begin{figure}[htb]
\includegraphics[width=\textwidth]{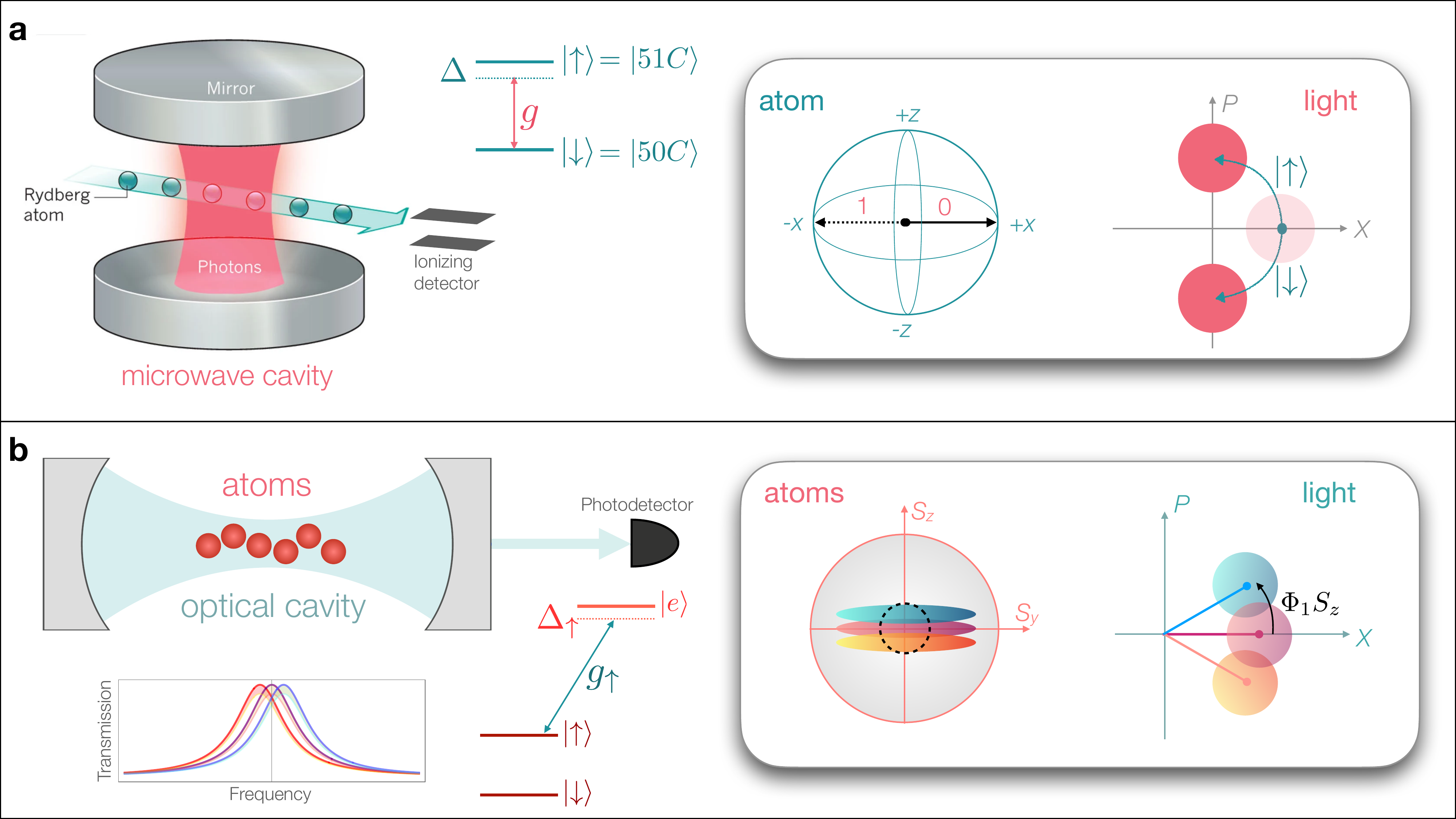}
\caption{\textbf{Quantum non-demolition (QND) measurements enabled by dispersive atom-light interactions.} (a) QND measurement of light using atoms.  Left: schematic, adapted from Ref.~\cite{hinds2012manipulating}, of microwave cavity probed by a beam of Rydberg atoms.  To probe the photon number parity, the transit time of an atom through a cavity is tuned to so that a single photon imparts a $\pi$ phase shift to the atom.  The backaction on the light is phase shift that depends on the atomic state, converting an initial coherent state of the light into a Schr\"{o}dinger cat state.  (b) QND measurement of atoms using light: the phase of light passing through the cavity becomes entangled with the collective spin component $S_z$, such that detecting the light enables a non-destructive measurement of $S_z$, projecting the atomic system into a state with squeezed quantum fluctuations. The backaction is an ac Stark shift that increases the uncertainty in $S_y$ due to photon shot noise in the light.}\label{fig:qnd_light_atoms}
\end{figure}

\subsection{QND Measurements of Light with Atoms as Probes}\label{sec:qnd_photons}

Quantum non-demolition measurements of light were demonstrated in pioneering experiments by S. Haroche~\cite{guerlin2007progressive,deleglise2008reconstruction} using the dispersive interaction of Eq.~\ref{eq:Hd}.  The essential concept, illustrated in Fig.~\ref{fig:qnd_light_atoms}(a), is to perform Ramsey spectroscopy with a single atom in a cavity to measure the phase $\Phi_n = 2n(g^2/\Delta)T$ acquired in a fixed time $T$ with $n$ intracavity photons.  For a time $T = \pi\Delta/(2g^2)$, corresponding to a single-photon phase shift $\Phi_1 = \pi$, the final atomic state is set by the parity (even or odd) of the photon number.  Thus, reading out the atomic state projects the light field into an eigenstate of photon number parity.  For a cavity mode initialized in a coherent state $\ket{\alpha}$, the resulting final state is a Schr\"odinger cat state $\ket{\psi} \propto \ket{i\alpha} \pm \ket{-i\alpha}$, as can be understood by considering the backaction of the measurement: the dispersive shift of the cavity mode frequency leads to a spin-dependent phase space precession by an angle $\pm\pi/2$ which, when reading out the spin in the $\uvec{x}$ basis, projects the light into a superposition of two coherent states separated in phase by $\Phi_1 = \pi$.

\begin{trailer}{Fundamental limit set by the cooperativity}
The QND measurement of light provides an instructive case study for the crucial role played by the cavity cooperativity.  In particular, let us ask how large a single-photon phase shift $\Phi_1$ we can achieve in an ideal setting where the atom has a stable ground state $\down$ and an excited state $\up$ of linewidth $\Gamma$.  If we want to avoid losing even one photon during a measurement with on average $\avg{n}$ photons, the interaction time $T$ is limited to $T \lesssim  1/(\avg{n}\tilde{\kappa})$, where
\begin{equation}\label{eq:tkappa}
\tkappa = \kappa + \left(\frac{g}{\Delta}\right)^2 \Gamma
\end{equation}
is the cavity loss rate, including both the intrinsic linewidth $\kappa$ and additional loss from atomic scattering.  Correspondingly, the single-photon phase shift is limited to
\begin{equation}
\Phi_1 \lesssim \frac{2g^2}{\Delta \avg{n} \tkappa} = \frac{2/\avg{n}}{\Delta \kappa/g^2 + \Gamma/\Delta}.
\end{equation}
The right-hand side is maximized at a detuning $\Delta = g \sqrt{\Gamma/\kappa} = \sqrt{\eta}\Gamma/2$, yielding
\begin{equation}
\Phi_1 \lesssim \frac{\sqrt{\eta}}{2\avg{n}}.
\end{equation}
We thus see that a large single-atom cooperativity $\eta \gtrsim 4\pi^2\avg{n}^2$ is essential to achieving the phase shift $\Phi_1 = \pi$ required for the QND measurement of parity.

\end{trailer}

The utility of the quantum non-demolition measurement of photon number goes even beyond the specific example of preparing Schr\"{o}dinger cat states.  The parity measurement is a tool not only for quantum control but also for quantum state characterization, enabling direct detection of the Wigner function~\cite{deleglise2008reconstruction} -- with odd photon parity signifying Wigner negativity, a key signature of nonclassicality.  Moreover, tuning the single-photon phase shift to values smaller than $\pi$ allows for counting higher photon numbers.  In Ref.~\cite{guerlin2007progressive}, by tuning the single-photon phase shift to $\Phi_1 = \pi/4$, letting a sequence of atoms individually transit the cavity, and measuring them in varying bases, the intracavity photon number was gradually collapsed from an initial coherent state into a Fock state.

These pioneering experiments with Rydberg atoms laid foundations for subsequent experiments with ``artificial atoms'' --- superconducting qubits --- where similar dispersive interactions with microwave resonators enable preparation of a wide range of non-classical states including Fock states~\cite{campagne2020quantum,eickbusch2022fast}, Schr\"{o}dinger cat states of 100 photons~\cite{vlastakis2013deterministically}, and Gottesman-Kitaev-Preskill (GKP) states for quantum error correction~\cite{campagne2020quantum,eickbusch2022fast}.

\subsection{QND Measurements of Atoms with Light as the Probe}\label{sec:qnd_atoms}

We now proceed to reverse the roles of atoms and light, using the atom-light entanglement engendered by the dispersive interaction of Eq.~\ref{eq:Hd} to prepare nonclassical \textit{atomic} states using \textit{light} as the probe.  One important application is the preparation of squeezed spin states as resources for enhanced precision measurement~\cite{schleier2010states,chen2011conditional,hosten2016measurement}, which has been demonstrated in ensembles of more than $10^5$ atoms~\cite{hosten2016measurement,bohnet2014reduced}.  In an opposite limit, a measurement of a light field that has interacted with two atoms has been used to prepare Bell pairs~\cite{welte2017cavity,dhordjevic2021entanglement}.  The unifying concept is to imprint information about the \textit{collective} atomic state onto a light field without revealing the states of individual atoms, such that a measurement of the light projects the atoms into an entangled state.  Beyond the application to engineering entanglement, a notable feature of cavity-aided measurement is that it can be fast and non-destructive, enabling mid-circuit measurements for quantum error correction~\cite{deist2022mid}.

\subsubsection{Squeezed Spin States}\label{sec:qnd_squeezing}

  A paradigmatic example of the application of QND measurement to engineering entanglement among atoms is the preparation of squeezed spin states for quantum metrology~\cite{schleier2010states}.  Such states are characterized by a reduction in quantum projection noise that enhances the sensitivity to rotations of a collective spin $\vec{S} = \sum_i \vec{s}_i$ formed by $N=2S$ two-level atoms.  The best angular resolution attainable without entanglement is the \textit{standard quantum limit} $\Delta\phi_\mathrm{SQL} = \Delta S_{\mathrm{CSS}}/S = 1/\sqrt{N}$, set by the quantum projection noise $\Delta S_\mathrm{CSS}$ of a coherent spin state (spin-polarized state).  Surpassing this limit requires introducing quantum correlations that reduce the noise while preserving coherence, as quantified by the Wineland parameter
  \begin{equation}\label{eq:wineland_parameter}
  \xi^2 \equiv N\frac{(\Delta S_\perp)^2}{\abs{\avg{\vec{S}}}^2},
  \end{equation}
  where $S_\perp$ denotes a spin component transverse to the mean spin vector $\avg{\vec{S}}$. A Wineland parameter $\xi^2 < 1$ signifies squeezing and implies an enhancement in spectroscopic sensitivity beyond the best achievable with the same $N$ particles in any unentangled state~\cite{wineland1994squeezed,pezze2018quantum}.
  
An ideal setup for cavity-based squeezing consists of an ensemble of $N$ atoms uniformly coupled to a cavity mode, such that the cavity couples to the collective spin $\vec{S}$, leading to a dispersive interaction
\begin{equation}\label{eq:Hd_coll}
H = \Odisp a^\dagger a S_z.
\end{equation}
Equation \ref{eq:Hd_coll} describes a shift of the cavity resonance frequency proportional to $S_z$, which enables a sensitive, nondestructive measurement of the collective spin; at the same time, it describes an ac Stark shift of the atomic transition that is proportional to the intracavity photon number, which will give rise to the measurement backaction.  In a conceptually simple scheme with two-level atoms detuned by $\Delta$ from cavity resonance, the coupling is $\Odisp = 2g^2/\Delta$ as in Sec.~\ref{sec:qnd_photons}.  In practice, a typical experimental scheme involves two (pseudo)-spin states $\ket{\downarrow},\ket{\uparrow}$ --- such as hyperfine~\cite{schleier2010states,chen2011conditional,bohnet2014reduced,hosten2016measurement}, Zeeman~\cite{braverman2019near,pedrozo2020entanglement}, and/or motional states~\cite{greve2022entanglement} --- which couple to the cavity via an optical transition to a third state $\ket{e}$ [Fig.~\ref{fig:qnd_light_atoms}(b)].  The interaction Hamiltonian then still takes the form in Eq.~\ref{eq:Hd_coll}, but with dispersive interaction $\Odisp = g_\uparrow^2/\Delta_\uparrow - g_\downarrow^2/\Delta_\downarrow$, where the atom-cavity coupling $g_\sigma$ and detuning $\Delta_\sigma$ depend on the state $\sigma\in \{\uparrow,\downarrow \}$.

The essential idea of squeezing by QND measurement is to first initialize the atoms in a spin-polarized state along $\uvec{x}$, then measure $S_z$ by optically probing the shift of the cavity resonance~[Fig.~\ref{fig:qnd_light_atoms}(b)].  Whereas a perfect projective measurement would prepare an eigenstate of $S_z$ (a Dicke state), a weaker measurement leaves the quantum system in a superposition of $S_z$ eigenstates with reduced variance $(\Delta S_z)^2 < (\Delta S_z)^2_\mathrm{CSS}$, compared with the variance $(\Delta S_z)^2_\mathrm{CSS} = N/4$ of the initial coherent spin state (CSS).  If this measurement is sufficiently nondestructive to preserve both $S_z$ and the transverse coherence $\avg{{S_x}}$, the effect is to project the atoms into an entangled state with reduced quantum fluctuations along $\uvec{z}$.  The backaction of the measurement of $S_z$ is an ac Stark shift that, due to photon shot noise in the probe light, imparts a fluctuating phase shift to the collective spin.  This backaction is essential to enforcing the Heisenberg uncertainty relation $\Delta S_y \Delta S_z \ge \abs{\avg{S_x}}$.

\begin{figure}[htb]
\includegraphics[width=\textwidth]{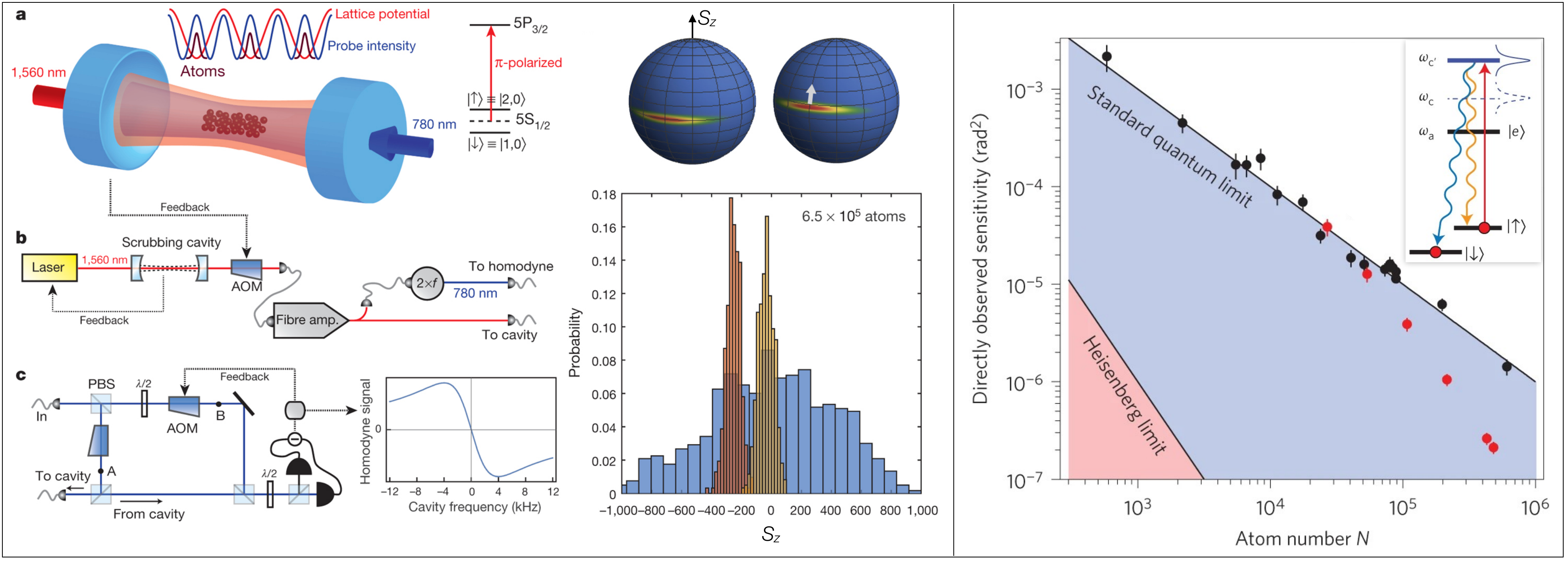}
\caption{\textbf{Spin squeezing by quantum nondemolition measurement.}  Left (\textit{adapted from Hosten et al.~\cite{hosten2016measurement}}): experimental setup used to achieve 20 dB of spin squeezing by cavity-aided QND measurement, with histograms of $S_z$ in the initial coherent spin state (blue) and in the squeezed state prepared by measurement, shown before and after the spin rotation indicated by the white arrow. Right (\textit{adapted from Bohnet et al.~\cite{bohnet2014reduced}}): spin squeezing on a cycling transition, where the Raman scattering process shown in blue is absent, yielding an angular sensitivity with Heisenberg scaling $(\Delta\phi)^2 \propto 1/N^2$.}\label{fig:qnd_squeezing}
\end{figure}

Experimental demonstrations of spin squeezing by cavity-aided QND measurement~\cite{schleier2010states,chen2011conditional,bohnet2014reduced,hosten2016measurement,cox2016deterministic,huang2023observing,robinson2024direct} have yielded gains of up to 20 dB ($\xi^{-2} = 100$) in angular sensitivity beyond the standard quantum limit~\cite{hosten2016measurement}.  The metrological applications of these states, including entanglement-enhanced clocks and atom interferometers, are discussed further in Sec.~\ref{sec:metrology}.

It is worth noting that squeezing generated by measurement is \textit{conditional}, in the sense that the particular state prepared depends on the measurement outcome $M$.  Nevertheless a squeezed state is prepared every time.  To make use of the conditional squeezing --- for example, in Ramsey spectroscopy --- it suffices to account for the conditional expectation value $\avg{S_z}_M$ in post-processing.  Conditional squeezing has also been converted to deterministic squeezing by applying classical feedback after the measurement~\cite{cox2016deterministic}, thereby always steering the system to the same state with $\avg{S_z}=0$.  In Ch.~\ref{ch:photon_mediated_interactions}, we will further see how a cavity enables deterministic squeezing via coherent feedback~\cite{schleier2010squeezing,leroux2010implementation,leroux2012unitary,zhang2015detuning}, which we will equivalently view as the cavity mediating interactions between the spins.

\begin{trailer}{Fundamental limits set by the cooperativity}

Spin squeezing requires operating in a regime of large collective cooperativity $N\eta \gg 1$.  This requirement arises because the measurement relies on information in the light scattered by the atoms into the cavity (at rate $\Gcav$).  This ``good'' scattering is necessarily accompanied by free-space scattering at a rate $\Gsc = \Gcav/(N\eta)$.  Each photon emitted into free space could in principle be traced back to a particular atom to learn its state, and thus reduces the collective spin coherence.  Quantifying this tradeoff will allow us to derive fundamental limits on spin squeezing set by the cooperativity.

For spin squeezing, it is convenient to operate in the regime where fluctuations in the spin projection $S_z$ due to quantum projection noise $\Delta S_z \propto \sqrt{N}$ shift the cavity resonance by less than a linewidth.  This condition, corresponding a dispersive coupling parameter $\Phi_1 < 1/\sqrt{N}$, facilitates obtaining a signal that depends linearly on $S_z$.  Specifically, an ideal measurement is performed with the probe tuned to cavity resonance for $S_z=0$, such that the phase of the output light depends linearly on $S_z$~\cite{hosten2016measurement}.  (Alternatively, for a detuned probe, the intensity of the output light also contains information about $S_z$~\cite{schleier2010states}.)  Conditioned on a measurement $M$ performed with on average $n$ intracavity photons, corresponding to a transmission rate $n\kappa/2$ for a two-sided cavity, the uncertainty in the collective spin projection is:
\begin{equation}
\Delta S_z|_M = \frac{\Delta\omega_c}{d\omega_c/dS_z} \propto \frac{\kappa/\sqrt{n\kappa t}}{\Odisp} \propto \frac{(n\kappa t)^{-1/2}}{\Phi_1}.
\end{equation}

It is instructive to reexpress the reduction in uncertainty in terms of the number of photons scattered into the cavity.  To this end, we use the relation $n\kappa\Phi_1^2 \propto \eta \Gsc$ between the cavity transmission rate $n\kappa$, the dispersive coupling $\Phi_1$, the free-space scattering rate $\Gsc$, and the cooperativity $\eta$.  The resulting variance in $S_z$, normalized to the projection noise level $(\Delta S_z)^2_\mathrm{CSS} = N/4$, depends on the measurement time $t$ as
\begin{equation}
\sigma^2_M \equiv \frac{(\Delta S_z)^2|_M}{N/2} \propto \frac{1}{N\eta\Gsc t}.
\end{equation}
Thus, the reduction in the variance of $S_z$ is set by the number of photons $N\eta\Gsc t$ forward-scattered into the cavity by the atoms, which improves with increasing collective cooperativity.

A fundamental limit on the squeezing is set by the degradation in spin coherence due to free-space scattering at rate $\Gsc t$.  In an ideal level scheme where the atoms are probed on a cycling transition, such that the only form of scattering is Rayleigh scattering, the optimal squeezing parameter $\xi^2 \sim \sigma^2 e^{2\Gsc t}$ is reached at a time $t_\mathrm{opt} \sim 1/\Gsc$, yielding $\xi^2_\mathrm{opt} \propto 1 /(N\eta)$~\cite{bohnet2014reduced} (Fig.~\ref{fig:qnd_squeezing}, right panel).  If the scattering also includes Raman processes that change the spin state, then we must additionally account for the noise added by Raman scattering, i.e., $\sigma^2_\mathrm{tot} \approx \sigma^2_M + r \Gsc t$, where $r$ is a constant factor set by the branching ratio for the scattering processes.  In this case, the noise is minimized at an earlier time $\Gsc t_\mathrm{opt} \propto 1/\sqrt{N\eta}$~\cite{sorensen2002entangling}.  In the regime $N\eta \gg 1$ required for strong squeezing, the reduction in spin length due to scattering is negligible at this early time $\Gsc t_\mathrm{opt} \ll 1$, so the squeezing parameter scales as $\xi^2 \approx \sigma^2 \propto 1/\sqrt{N\eta}$.
\end{trailer}

\subsubsection{Non-Gaussian W States}

\begin{figure}[htb]
\includegraphics[width=\textwidth]{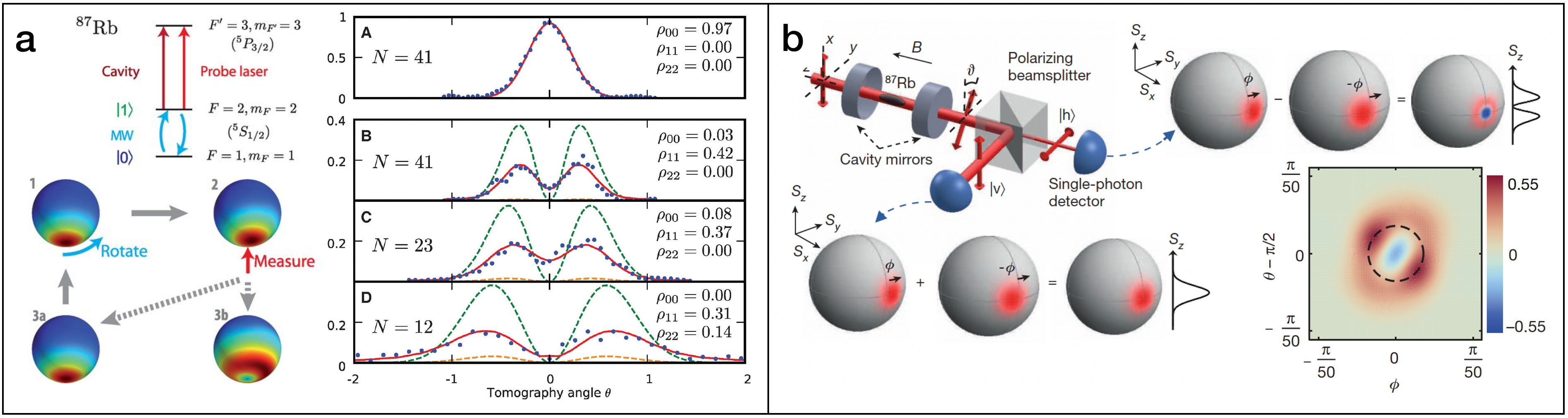}
\caption{\textbf{Preparation of W states by measurement:} (a) \textit{Adapted from Ref.~\cite{haas2014entangled}.}  Preparation of a W state in a strong-coupling cavity by by probing the vacuum Rabi splitting.  (b) \textit{Adapted from Ref.~\cite{mcconnell2015entanglement}.} Heralded preparation of a W state at weak single-atom cooperativity via Faraday interaction and single-photon detection.}\label{fig:W_states}
\end{figure}

Quantum nondemolition measurements have also been applied to prepare non-Gaussian collective spin states.  An illuminating example is provided by two experiments showing complementary measurement-based protocols for preparing a W state, consisting of a single collective excitation delocalized over a system of $N$ atoms.  One approach, demonstrated by Haas \textit{et al.}~\cite{haas2014entangled} and illustrated in Fig.~\ref{fig:W_states}, requires operating deep in the single-atom strong coupling regime $\eta \gg 1$ and leverages the vacuum Rabi splitting to detect the presence of an atom in state $\ket{\uparrow}$.  Specifically, the cavity is tuned to resonance with an optical transition $\ket{\uparrow}\rightarrow\ket{e}$, and the atoms are initially all prepared in state $\ket{\downarrow}$, where the pseudo-spin is encoded in a pair of hyperfine states.  A weak microwave pulse rotates the atomic state, producing a small amplitude for an atom to be in state $\ket{\uparrow}$.  For a light field tuned to be on resonance if all atoms are in state $\ket{\downarrow}$, the vacuum Rabi splitting suppresses transmission if a single excitation has been created.  Thus, the absence of transmission heralds successful creation of the $W$ state.  This approach was used to prepare demonstrably entangled $W$ states of up to 41 atoms in fiber cavity with cooperativity $\eta \sim 10^2$.

An alternative approach accessible even at weak single-atom cooperativity $\eta < 1$ was proposed and demonstrated in the group of V. Vuletic~\cite{mcconnell2013generating,mcconnell2015entanglement}.  Here, an ensemble of $N=3\times 10^3$ atoms with spins encoded in Zeeman states was prepared in a spin-polarized state along the $\uvec{x}$ axis and coupled to light in a cavity via the Faraday interaction~\cite{hammerer2010quantum} [Fig.~\ref{fig:W_states}(b)]
\begin{equation}
H_F = \Odisp \left(a_+^\dagger a_+ - a_-^\dagger a_- \right)S_z.
\end{equation}
Here, $a_\pm = H \pm iV$ represent annihilation operators for the two circular polarizations $\sigma_\pm$ supported by the cavity, such that $H_F$ describes a vector ac Stark shift acting on the atomic Zeeman states or, equivalently, an $S_z$-dependent polarization rotation.  A polarizer at the cavity output is tuned such that the output port is dark for $S_z =0$, whereas a small nonzero value $\abs{S_z}>0$ rotates the polarization.  For a weak coherent pulse of light incident on the cavity, a photodetection event then heralds the preparation of a state with nonzero $S_z$, and crucially does so without revealing the sign of $S_z$.  Conditioned on detecting a photon, the final $S_z$ distribution is thus bimodal.  Moreover, the measurement backaction affects the atomic phase: since the detected photon is in a superposition of two circular polarizations that rotate the spin in opposite directions about $\uvec{z}$, the final atomic state can equivalently be viewed as a superposition of two coherent states separated by a phase $\Phi = \Odisp/\kappa$.  In the limit where this phase is much smaller than the projection noise of the coherent spin state, $\Phi \ll 1/\sqrt{N}$, the result is a $W$ state with a single excitation in the basis of $S_x$ eigenstates~[Fig.~\ref{fig:W_states}(b)].  In principle, for sufficiently strong atom-light coupling, the same protocol also enables preparation of well separated superpositions of coherent states, i.e., Schr\"{o}dinger ``kitten'' ($\Phi > 1/\sqrt{N}$) or cat ($\Phi = \pi$) states~\cite{mcconnell2015entanglement}, and generalizes to more versatile quantum control (Sec.~\ref{sec:carving_painting}).

The experiments in Refs.~\cite{haas2014entangled} and \cite{mcconnell2015entanglement} are similar in enabling heralded preparation of W states, but a key difference lies in the requirement on the cavity cooperativity.  For the method of Ref.~\cite{haas2014entangled} employing the vacuum Rabi splitting, strong coupling $\eta \gg 1$ is essential (and, additionally, the conditions $g\gg \kappa,\Gamma$ must be independently satisfied, motivating the use of a short cavity to achieve $g \gg \Gamma$).  For the method employing the Faraday interaction and heralding on detection of a single photon, strong coupling is still advantageous but can be traded off with detection efficiency~\cite{mcconnell2013generating}.  Thus, even at a cooperativity $\eta \approx 0.1$ in Ref.~\cite{mcconnell2015entanglement}, it was possible to prepare a $W$ state with high purity, enabling the observation of Wigner negativity and a provable entanglement depth of 2910(190) atoms at a total atom number $N=3100$.

\subsubsection{Towards Full Quantum Control: Carving or Painting}\label{sec:carving_painting}

\begin{figure}[htb]
\includegraphics[width=\textwidth]{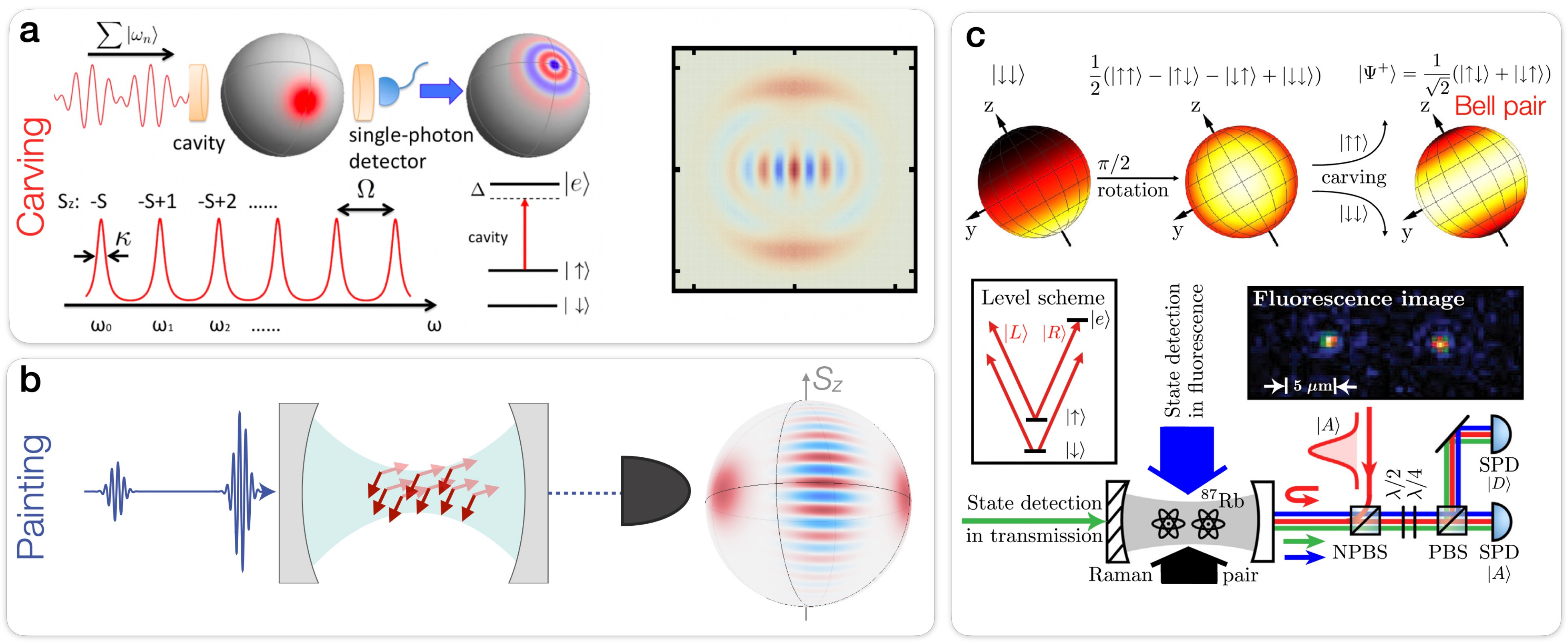}
\caption{\textbf{Carving or painting entangled states via QND measurement with shaped single photons.} (a-b) Theoretical proposals from Refs.~\cite{chen2015carving,davis2018painting} for quantum control of collective spin states. (a) The frequency spectrum of the drive field determines which Dicke states are ``carved'' out of the initial superposition state.  (b) Equivalently, the intensity waveform of the drive field paints a desired superposition of coherent states on the collective Bloch sphere.  An exponential pulse envelope ensures that the arrival time of the photon at the detector carries no information about the time of incidence on the cavity. (c) Experimental implementation of two-qubit entanglement by carving, from Ref.~\cite{welte2017cavity}.}\label{fig:carving_painting}
\end{figure}

A promising future prospect is to leverage heralded state preparation based on single-photon detection for more flexible quantum control.  A versatile scheme proposed in Refs.~\cite{chen2015carving,davis2018painting,ramette2025carving} is to engineer the frequency spectrum --- or, equivalently the pulse shape --- of a weak coherent field incident on a cavity, containing on average less than one photon.  When the $S_z$-dependent cavity shift $\Odisp > \kappa$ is large enough to resolve individual Dicke states, this method allows for preparing arbitrary superpositions of Dicke states specified by the frequency components of the drive field.  Equivalently, in a time-domain picture, for a short pulse incident on the cavity, detection of a photon a time $T$ later heralds the preparation of a coherent state whose phase has been rotated by an angle $\Phi_T = \Odisp T$. Thus, an incident field comprised of two short pulses prepares a superposition of two coherent states, i.e., a Schr\"{o}dinger kitten or cat state (\ref{fig:carving_painting}).

More generally, we may view the drive field as ``painting'' an arbitrary superposition of coherent states on the Bloch sphere~\cite{davis2018painting}.  For example, a continuous pulse of duration $2\pi/\Odisp$ theoretically paints a circle on the Bloch sphere, producing a Dicke state.  Scaling the pulse amplitude with an exponentially growing envelope ensures that the arrival time of the photon contains no information that would decohere the superposition of coherent states~\cite{davis2018painting}.

The time-domain picture again highlights the value of strong dispersive coupling: full quantum control generically requires imparting phase shifts as large as $\Phi = \Odisp T = 2\pi$.  To achieve a moderate success probability in the heralded scheme, this phase shift should be imparted in a time $T\lesssim1/\tkappa_N$, where $\tkappa_N = \kappa + N\Gsc$ is the photon loss rate including the effect of atomic scattering. The ratio $\tkappa/\Odisp$ is minimized at an optimal detuning $\Delta \propto \sqrt{N\eta} \Gamma$, yielding a single-photon phase shift $\Phi_1 \propto \sqrt{\eta/N}$.  Larger phase shifts $\Phi = \Phi_1 \kappa T > \sqrt{\eta/N}$ can be achieved only at the cost of an exponentially decaying success probability.  Thus, roughly, a single-atom cooperativity $\eta > 1$ is required to prepare a ``kitten'' state comprising a well-separated superposition of coherent states, while $\eta > N$ is required to access a maximally macroscopic superposition (Schr\"{o}dinger's cat) state.

To date, the method of carving has been applied in systems of $N=2$ atoms to generate entangled Bell pairs~\cite{welte2017cavity,dhordjevic2021entanglement}.  The approach has been applied both in a macroscopic cavity~\cite{welte2017cavity} and in a nanophotonic platform~\cite{dhordjevic2021entanglement}, yielding fidelities in the range of 70-90\%. Although these fidelities are not competitive with those attainable using local (Rydberg blockade) gates, the ability to generate long-range, photon-mediated entanglement is nevertheless of interest for quantum networking, e.g., to enable modular computing architectures employing optical interconnects~\cite{ramette2022any,ramette2024fault}.

\subsubsection{Cavity-Aided Measurement of Single Atoms}

Cavity-aided nondemolition measurements in arrays of single atoms open new opportunities for quantum error correction and quantum networking~\cite{ramette2022any,ramette2024fault}.  In particular, a high-cooperativity cavity enables fast, nondestructive qubit state discrimination by channeling the light scattered by a single atom into a single mode to enable high detection efficiency~\cite{bochmann2010lossless,gallego2018strong,gehr2010cavity,dhordjevic2021entanglement,hu2025site,shaw2025cavity}.  An experiment by Deist \textit{et al.}~\cite{deist2022mid} highlighted this capability in a minimal setting of two atoms, each of which could be independently moved into and out of the cavity by an optical tweezer.  The hyperfine state of an atom in the cavity was detected in 25~\micro{s} with fidelity above 99\% while leaving a superposition state of the other atom unperturbed, thereby demonstrating a coherence-preserving mid-circuit measurement.  Recent experiments have also demonstrated site-selective readout based on local optical addressing~\cite{hu2025site} and developed arrays of cavities for efficient parallel readout of atom arrays~\cite{shaw2025cavity}.

\section{Photon-mediated interactions}\label{ch:photon_mediated_interactions}

By channeling information about the collective state of two or more atoms into a single mode of light without revealing the states of individual atoms, a cavity is naturally suited to generating entanglement.  Whereas we so far considered the case where we detect the light to project the atoms into an entangled state, we may alternatively induce the light to act back on the atoms in a manner that depends on the atomic states, thereby mediating interactions among the atoms.  Beyond providing a complementary approach to engineering entangled states, such photon-mediated interactions are of interest for quantum simulations, notably providing access to models with nonlocal connectivity.

In this section, we will first build intuition for how the \textit{atom-light} interactions examined in Secs.~\ref{sec:introduction}-\ref{sec:qnd_measurement} give rise to \textit{atom-atom} interactions \textit{mediated} by light.  We will start from examples of collective spin models with applications in quantum metrology, then introduce emerging tools for venturing beyond single-mode dynamics into richer many-body physics.  

\subsection{Collective Spin Dynamics}\label{sec:collective_spin}

\subsubsection{Ising interactions}\label{sec:Ising}
Cavity-mediated spin-spin interactions were first demonstrated in the context of quantum metrology.  A driving motivation was a seminal 1993 paper by Kitagawa and Ueda that proposed the \textit{one-axis twisting} Hamiltonian
\begin{equation}\label{eq:Htwist}
\Htwist = \chi S_z^2/N,
\end{equation}
consisting of collective Ising interactions, as a mechanism for spin squeezing~\cite{kitagawa1993squeezed}.  The authors included a disclaimer that ``realistic physical schemes [for implementing $\Htwist$] are yet to be found,'' highlighting how exotic the requisite all-to-all interactions seemed at the time.

\begin{figure}[htb]
\includegraphics[width=\textwidth]{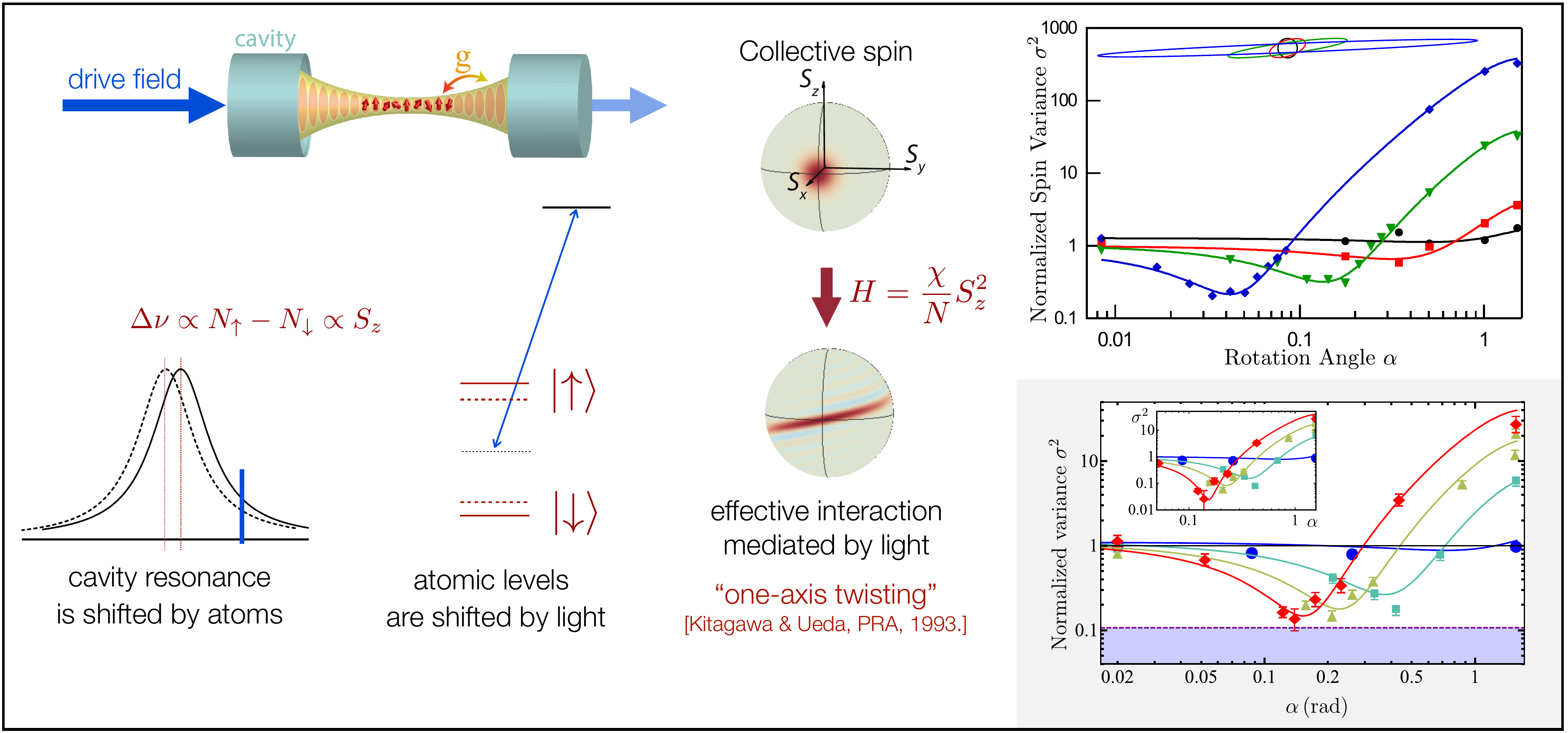}
\caption{\textbf{One-axis twisting via cavity-mediated Ising interactions.} For a drive field detuned from cavity resonance, the $S_z$-dependent cavity resonance shift produces an $S_z$-dependent intracavity intensity which acts back on the atoms via the ac Stark shift to induce an $S_z$-dependent spin precession rate, twisting the Bloch sphere about the $\vec{z}$ axis.  Data figures, reprinted from Ref.~\cite{leroux2010implementation} (top) and Ref.~\cite{braverman2019near} (inset), show spin variance vs rotation angle for the resulting squeezed states.  Top: squeezing in hyperfine clock states of $^{87}$Rb.  The one-axis twisting at normalized drive detuning $d=1$ is dissipative, yielding antisqueezing significantly stronger than the squeezing.  Inset: near-unitary squeezing for an optimized driving scheme on an almost closed transition in $^{171}$Yb~\cite{braverman2019near}.}\label{fig:twisting}
\end{figure}

By now, several experiments have realized cavity-mediated Ising interactions (one-axis twisting) and the resulting spin squeezing.  The prevailing approach relies on the same dispersive interaction used for spin squeezing by QND measurement (Eq.~\ref{eq:Hd_coll}).  The key idea is to drive the cavity with a field that is detuned from resonance, such that the $S_z$-dependent shift of the cavity resonance produces an $S_z$-dependent intracavity intensity.  The resulting linear dependence of intracavity photon number on $S_z$ produces an $S_z$-dependent spin precession, as shown in Fig.~\ref{fig:twisting}.  For a system initialized in a spin-polarized state along $\uvec{x}$, the effect of this twisting is to squeeze the quantum fluctuations in the $yz$-plane.  In Sec.~\ref{sec:metrology}, we describe applications of the resulting squeezing in entanglement-enhanced spectroscopy.

To arrive at the effective spin-spin interaction Hamiltonian $\Htwist$, we express the intracavity photon number for a driven cavity in terms of $S_z$.  The mean steady-state intracavity photon number is a Lorentzian function $\avg{a^\dagger a} \propto L\left(\frac{\delta + \Odisp S_z}{\kappa/2}\right)$ of the $S_z$-dependent detuning, where $\delta$ is the drive-cavity detuning at $S_z=0$ and
$L(u) = 1/(1+u^2)$.  Assuming that the spin-dependent shift of the cavity resonance is kept small compared to the drive detuning, $\Odisp S_z \ll \delta$, expanding the Lorentzian about $S_z = 0$ yields to lowest order a linear dependence
\begin{equation}
\avg{a^\dagger a} \approx n_0\left(1 + \frac{2d}{1+d^2}\frac{2\Odisp}{\kappa}S_z\right)
\end{equation}
of intracavity photon number on $S_z$, where $d = 2\delta/\kappa$ and $n_0 = \avg{a^\dagger a}_{S_z=0}$.  The result is an effective Hamiltonian~\cite{davis2016approaching}
\begin{equation}\label{eq:driven_cavity_twisting}
H_\mathrm{eff} \approx n_0\Odisp S_z + J S_z^2,
\end{equation}
where the second term describes Ising interactions of strength
\begin{equation}
J = \frac{4n_0\Odisp^2}{\kappa}\frac{d}{1+d^2}
\end{equation}
among all atom pairs.  The first term in Eq.~\ref{eq:driven_cavity_twisting} is an effective field due to the average ac Stark shift, which is readily removed by spin echo.  We are thus left with the one-axis twisting Hamiltonian $\Htwist = \chi S_z^2/N$.  For large drive detuning $\delta \gg \kappa$, the pairwise interaction strength $J = \chi/N$ reduces to
\begin{equation}\label{eq:chi_twist}
J \approx \frac{2 n_0 \Omega^2}{\delta} = \frac{2 n_0 g^4}{\Delta^2 \delta} = \eta \Gsc \frac{\kappa}{2\delta}.
\end{equation}
On the right-hand side, by re-expressing the interaction strength in terms of the free-space scattering rate $\Gsc$, we can already begin to see how the cooperativity $\eta$ will govern the coherence of the interactions.

\begin{trailer}{Coherence of photon-mediated interactions}

Two effects fundamentally limit the coherence of photon-mediated interactions.  In the limit of large drive detuning $\delta$ shown in Eq.~\ref{eq:chi_twist}, free-space scattering becomes the dominant limitation, since the interaction strength at fixed intracavity photon number decreases as $1/\delta$.  However, in the opposite limit of a drive field near cavity resonance, the dominant limitation is that the outgoing light carries away information about the collective spin state --- the very same information that enables spin squeezing by measurement (Sec.~\ref{sec:qnd_squeezing}).  The backaction of the measurement is a fluctuating ac Stark shift due to photon shot noise in the intracavity light, which introduces collective dephasing of the atomic spins, formally described by a Lindblad operator $\mathcal{L} = \sqrt{\gamma} S_z$ with dephasing rate $\gamma = J\kappa/\delta$~\cite{davis2016approaching}. The scaling $\gamma/J \propto 1/\delta$ can be understood from the fact that, for a fixed interaction strength $J$, operating at larger detuning requires a higher incident photon rate, such that the photon shot noise is fractionally smaller.  At finite cooperativity, choosing an optimal detuning $\delta\sim\sqrt{\eta}\kappa$ maximizes the interaction-to-decay ratio $J/(\gamma+\Gsc)\sim \sqrt{\eta}$ for pairwise interactions~\cite{zhang2015detuning,davis2016approaching}.

For the application to spin squeezing by one-axis twisting in an ensemble of $N$ atoms, the \textit{collective} cooperativity $N\eta$ imposes limits similar to those encountered in measurement-based squeezing.  The geometry of the squeezed state (Fig.~\ref{fig:twisting_geometry}) is governed by the twisting strength $Q = NJt = N\eta \Gsc t / d$, where $d \equiv 2\delta/\kappa$ is the dimensionless drive-cavity detuning.   In the limit $Q\gg 1$ required for strong squeezing, and in an ideal level scheme with Rayleigh scattering only, the squeezed quadrature shrinks as $\sigma^2 \propto 1/Q^2 + 1/Qd \gtrsim 1/(N\eta\Gsc t)$, permitting a squeezing parameter $\xi^2 \propto 1/(N\eta)$ at $\Gsc t \sim 1$.  Two effects that may thwart this scaling are non-Gaussianity arising from the curvature of the Bloch sphere; and noise added by Raman scattering.  The curvature of the Bloch sphere imposes a limitation $\xi^2 \propto N^{-2/3}$ that applies even to the ideal, unitary one-axis twisting Hamiltonian~\cite{kitagawa1993squeezed}.  (A protocol for accessing the full metrological benefit of the non-Gaussian state in this regime is presented in Sec.~\ref{sec:time_reversal}).  Raman scattering adds noise $\sigma^2_\mathrm{sc} \propto \Gsc t = Qd/(N\eta)$, limiting the squeezing to $\xi^2 \propto 1/\sqrt{N\eta}$.  This limit was first derived by S{\o}rensen and M{\o}lmer for an alternative implementation of cavity-mediated one-axis twisting in Ref.~\cite{sorensen2002entangling}.

\begin{figure}[htb]
\centering
\includegraphics[width=0.5\textwidth]{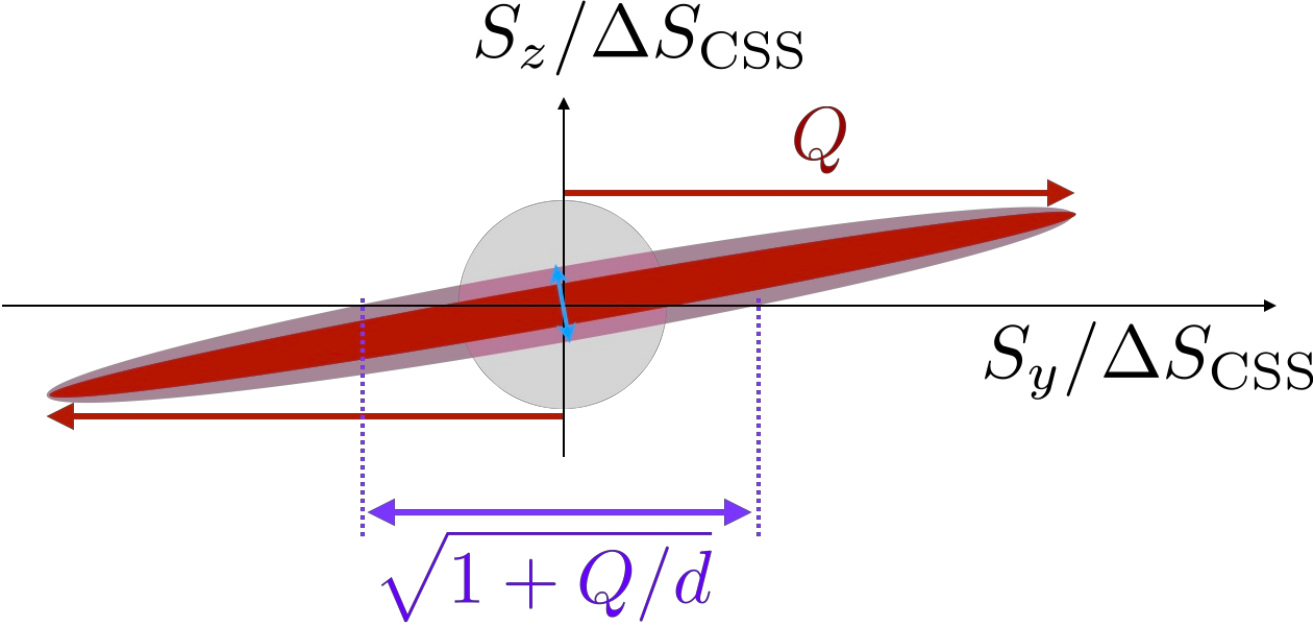}
\caption{\textbf{Geometry of the squeezed state prepared by cavity-mediated one-axis twisting,} in terms of the twisting strength $Q = NJt$ and normalized detuning $d=2\delta/\kappa$.  Red ellipse shows ideal unitary evolution; purple ellipse includes broadening due to collective dephasing at finite detuning $d$; gray circle shows coherent spin state for reference.}\label{fig:twisting_geometry}
\end{figure}
\end{trailer}

\subsubsection{Spin-exchange interactions}\label{sec:spin_exchange}

\begin{figure}[htb]
\includegraphics[width=\textwidth]{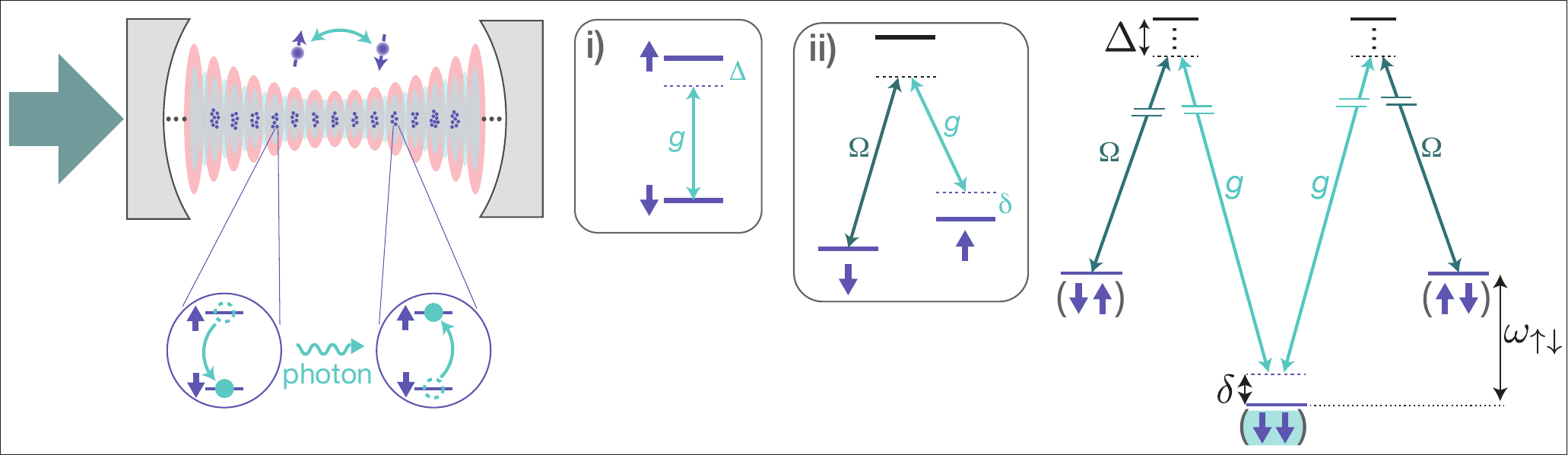}
\caption{\textbf{Cavity-mediated spin exchange interactions.}  (i) Level scheme for direct coupling of two-level atoms to a cavity.  (ii) Level scheme for coupling two ground states to an optical cavity via a Raman transition controlled by a drive field of Rabi frequency $\Omega$.  Right: two-atom level scheme showing the virtual process in which an atom flips its spin by scattering a photon into the cavity. Reabsorption of this photon by a second atom completes the spin exchange process.}
\label{fig:spin_exchange}       %
\end{figure}

Perhaps the conceptually simplest form of photon-mediated interaction is the cavity-mediated spin exchange shown in Fig.~\ref{fig:spin_exchange}, where two atoms interact via virtual exchange of a cavity photon.  For a system in initialized in state $\ket{\uparrow\downarrow}\otimes\ket{0}_c$, the first atom can emit a photon into the cavity to access state $\ket{\downarrow\downarrow}\otimes\ket{1}_c$, and subsequent reabsorption of the photon by the second atom yields $\ket{\downarrow\uparrow}\otimes\ket{0}_c$. For large atom-cavity detuning $\delta = \omega_0 - \omega_c \gg g$, the intermediate state is only virtually populated and mediates resonant ``flip-flop'' processes between the atoms, with spin-exchange coupling $J\sim g^2/\delta$.  Formally, the underlying atom-light interactions are described by the Tavis-Cummings Hamiltonian, which generalizes the Jaynes-Cummings model of Eq.~\ref{eq:Hjc} to a system of $N$ atoms:
\begin{equation}\label{eq:Htc}
\Htc = g \left(a^\dagger S_- + a S^+\right) + \Delta S_z,
\end{equation}
where $\vec{S} = \sum_{i=1}^N\vec{s}_i$ denotes the collective spin.  For large detuning $\Delta \gg g\sqrt{N}$, the cavity mode can be adiabatically eliminated from the dynamics, leaving an effective spin-spin interaction
\begin{equation}\label{eq:Hflipflop}
H = \frac{g^2}{\Delta} S_+ S_-.
\end{equation}

Collective spin-exchange interactions have been observed in ensembles of strontium atoms by coupling a cavity to the optical clock transition~\cite{norcia2018cavity}, and in alkali atoms by coupling two stable ground states to a cavity via Raman transitions~\cite{davis2019photon,davis2020protecting,periwal2021programmable} [Fig.~\ref{fig:spin_exchange}(ii)].  In the latter scenario, the Hamiltonian takes the general form
\begin{equation}\label{eq:HflipflopR}
H_{XY} = \frac{\chi_- S_+ S_- + \chi_+ S_- S_+}{N}.
\end{equation}
Here, $\chi_\pm = N\mathcal{G}^2/\delta_\pm$, where $\delta_\pm = \omega_d - \omega_c \pm \omega_{\uparrow\downarrow}$ are the detunings of two possible Raman processes in which an atom flips its spin up/down by absorbing a photon from a drive field and emitting into the cavity; and $\mathcal{G} = g\Omega/(2\Delta)$ is the two-photon Rabi frequency of these Raman processes in terms of the Rabi frequency $\Omega$ of the drive field and detuning $\Delta$ of the cavity from the optical transition.  For large drive-cavity detuning $\delta_\pm \gg \omega_{\uparrow\downarrow}$, the ``flip-flop'' and ``flop-flip'' couplings $\chi_\pm$ become approximately equal.  Further, in the case where the drive field is incident through the cavity, the Rabi frequency $\Omega = 2\sqrt{n_0}g$ for $n_0$ photons in the drive mode yields $J =\chi_\pm/N \sim n_0 g^4/(\Delta^2 \delta)$, paralleling the Ising coupling in Eq.~\ref{eq:chi_twist}.

While direct spin-exchange on a long-lived optical transition is relevant for prospective applications in entanglement-enhanced clocks, the Raman scheme offers benefits including dynamical control of the interactions via a drive field, flexibility for engineering the form~\cite{davis2020protecting,luo2025hamiltonian} and spatial structure~\cite{hung2016quantum,periwal2021programmable} of the spin-spin couplings, and generalizations to multibody interactions~\cite{luo2024realization}.  While one natural encoding for the Raman scheme is in Zeeman states~\cite{davis2020protecting,periwal2021programmable}, recent work has also explored the use of momentum states as the pseudo-spin states~\cite{luo2025hamiltonian,finger2024spin}, with prospective applications in entanglement-enhanced matter-wave interferometry.  Both of these encodings offer the convenient feature that the frequency splitting $\omega_{\uparrow\downarrow}$ can be comparable to the cavity linewidth, allowing drive fields detuned by $\delta_\pm \gg \omega_{\uparrow\downarrow}$ to be incident through the cavity.

Experiments realizing cavity-mediated spin-exchange interactions have probed the mean-field dynamics of the collective spin~\cite{norcia2018cavity}, imaged excitation transfer in spatially extended systems~\cite{davis2019photon,davis2020protecting}, and observed signatures of the formation of correlated atom pairs in ensembles of spin-1 atoms~\cite{davis2019photon,periwal2021programmable,cooper2024graph,finger2024spin}, including spin-nematic squeezing~\cite{masson2017cavity,cooper2024graph} (Sec.~\ref{sec:spin_mixing}).  Generalizations incorporating bichromatic or multichromatic driving allow for engineering XYZ models that include ``flip-flip'' and ``flop-flop'' terms~\cite{sorensen2002entangling,borregaard2017one,mivehvar2019cavity}, enabling recent realizations of two seminal theoretical models~\cite{luo2024realization}: the M{\o}lmer-S{\o}rensen interaction $H\propto S_x^2$ ~\cite{sorensen2002entangling} and the two-axis countertwisting Hamiltonian $H_\mathrm{ct} \propto S_x^2 - S_y^2$~\cite{kitagawa1993squeezed}.  Multichromatic driving further provides control over the spatial structure of interactions, as described in Sec.~\ref{sec:programmable}.

In systems with only global control and detection, the spin-exchange Hamiltonian is nearly equivalent to the one-axis twisting Hamiltonian of Eq.~\ref{eq:Htwist}.  In particular, Eq.~\ref{eq:HflipflopR} can be reexpressed as $H_{XY} \propto S(S+1) - S_z^2$.  For a system initialized in an eigenstate of total angular momentum $S$, as in the common case of a spin-polarized initial state, only the ``twisting'' term $S_z^2$ induces dynamics. However, a key difference from pure one-axis twisting $\Htwist$ lies in the robustness to perturbations: the $\abs{\vec{S}}^2$ term added by spin-exchange interactions produces an energy gap between manifolds of different total spin $S$~\cite{norcia2018cavity}, which serves to protect the collective spin coherence~\cite{davis2020protecting}.

\subsubsection{Tunable Heisenberg Interactions}\label{sec:tunable_heisenberg}

At first glance, the mechanisms underlying cavity-mediated Ising interactions (Sec.~\ref{sec:Ising}) and spin-exchange interactions (Sec.~\ref{sec:spin_exchange}) might seem quite distinct.  We described the former in terms of a spin-dependent ac Stark shift and the latter in terms of scattering processes --- yet, in fact, these are complementary descriptions of essentially the same physics, and either picture can be used to describe both forms of interaction.  The equivalence builds on the relationship between Raman scattering and ac Stark shifts: a Raman coupling between Zeeman states generically arises from a modulated vector light shift that acts as an effective transverse field to drive spin flips.  Spin-exchange interactions mediated by a cavity can be understood similarly: in the presence of a magnetic field perpendicular to the cavity, the spins precess and thereby modulate the circular birefringence of the cavity via the Faraday effect, producing an oscillating vector light shift that drives spin flips.

\begin{figure}[htb]
\includegraphics[width=\textwidth]{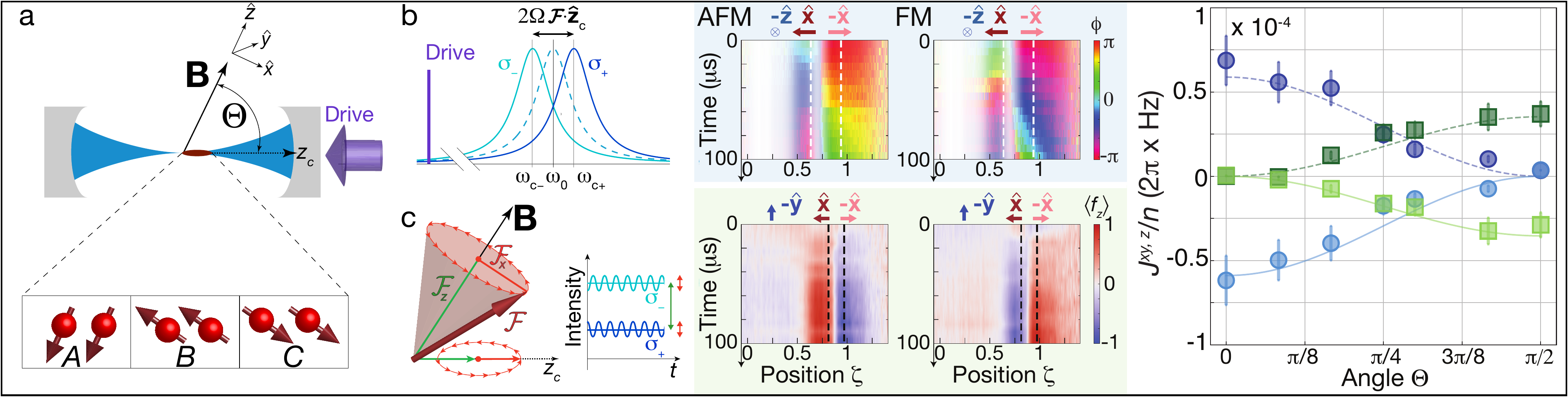}
\caption{\textbf{Tunable Heisenberg interactions}. \textit{Figure adapted from Ref.~\cite{davis2020protecting}.} Left: Scheme for tuning the Heisenberg couplings $J^{xy,z}$ via the angle $\theta$ of a magnetic field from the cavity axis.  Center: mean-field dynamics of a spatially extended atomic ensemble under Ising (top) and spin-exchange (bottom) interactions, used to verify the dependence of the Ising coupling $J^z$ (blue) and spin-exchange coupling $J^{xy}$ (green) on angle.  The choice of ferromagnetic (FM) or antiferromagnetic (AFM) couplings is governed by the sign of the drive-cavity detuning.}
\end{figure}

The equivalence of the light-shift and scattering pictures becomes most evident when we continuously tune between Ising couplings and spin-exchange couplings~\cite{bentsen2019integrable,davis2020protecting}.  In the experiment of Ref.~\cite{davis2020protecting}, with spins encoded in Zeeman states, we demonstrated that generic Heisenberg XXZ models
\begin{equation}\label{eq:Hxxz}
H_\mathrm{XXZ} = J_{xy}(S_x^2 + S_y^2) + J_z S_z^2
\end{equation}
with tunable anisotropy $J_z/J_{xy}$ can be realized by tuning the orientation of a magnetic field relative to the cavity axis.  We applied this tunability to directly compare the robustness of the collective spin to an inhomogeneous field in the presence of pure Ising or pure spin-exchange interactions, observing that spin-exchange interactions protect the spin coherence.  In a generalization to inhomogeneous atom-cavity couplings, theoretical work predicts a rich dynamical phase diagram of the tunable XXZ model, discussed further in Sec.~\ref{sec:random_spin_models}.

\subsubsection{Higher Spin: Cavity-Mediated Spin Mixing}\label{sec:spin_mixing}

\begin{figure}[htb]
\includegraphics[width=\textwidth]{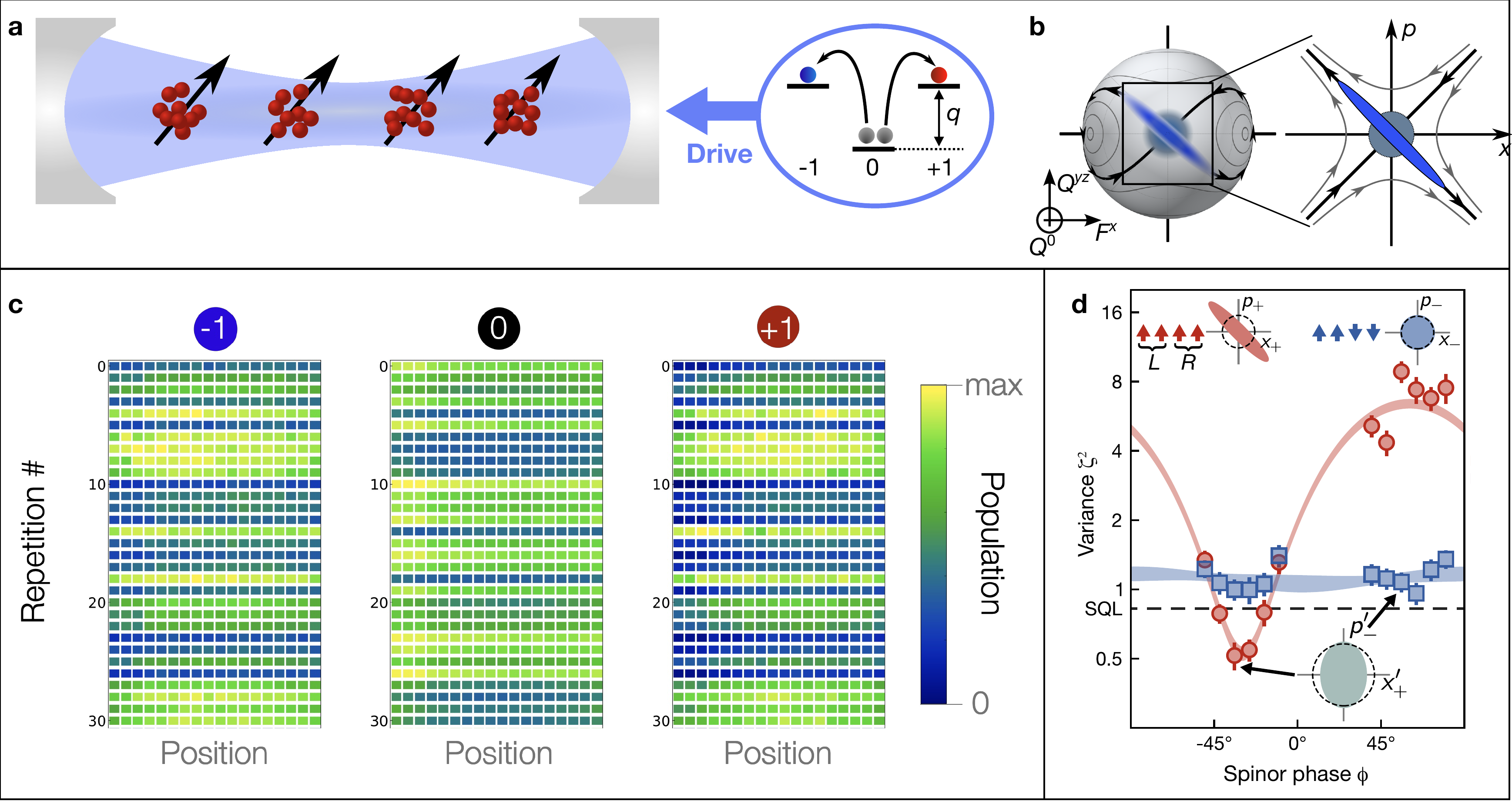}
\caption{\textbf{Cavity-mediated spin mixing} in an array of ensembles of spin-1 atoms~\cite{davis2019photon,periwal2021programmable,cooper2024graph}. (a) Experimental setup for observing cavity-mediated spin nematic squeezing in Ref.~\cite{cooper2024graph}. (b) Squeezing dynamics on a generalized Bloch sphere spanned by $(F_x, Q^{yz},Q^0)$, with mapping to quadrature operators $x\propto F^x$, $p\propto Q^{yz}$. (c) Observation of cavity-mediated spin mixing in an array of 18 atom clouds initialized in $m=0$, showing macroscopic fluctuations in population of $m=\pm 1$ that are correlated across the millimeter-scale array. (d) Normalized spin variance vs spinor rotation angle, showing spin nematic squeezing. \textit{Subfigures (a), (b), (d) are adapted from Ref.~\cite{cooper2024graph}.  Subfigure (b) shows representative raw data for measurements reported in Ref.~\cite{periwal2021programmable}.}}\label{fig:spin_mixing}
\end{figure}

Our description of cavity-mediated spin-exchange as arising from the Faraday interaction between the atomic magnetization and light --- a seemingly classical effect --- would be incomplete without discussing the role of quantum fluctuations.  Notably, even an atomic system with no \textit{average} magnetization can couple to the cavity and undergo dynamics.  A striking signature arises in systems of spin-1 atoms initialized in the $m_f = 0$ state, where spin-exchange couplings enable the formation of correlated atom pairs in $m_f = \pm 1$ (Fig.~\ref{fig:spin_mixing}).  This process was proposed theoretically as a mechanism for cavity-mediated spin-nematic squeezing~\cite{masson2017cavity} and has been observed in Refs.~\cite{davis2019photon,periwal2021programmable,cooper2024graph}.

\subsubsection{Dicke Model and Self-Organization}
In our discussion of photon-mediated interactions, we so far focused primarily on the Ising and spin-exchange cases, where the number of spin excitations is (ideally) conserved.  Formally, this conservation can be traced back to two implicit assumptions: (1) the use of the rotating wave approximation to arrive at the Jaynes-Cummings Hamiltonian, which conserves the total number of spin and photonic excitations; and (2) the assumption that the field scattered by the atoms into the cavity is weak, which allows for adiabatically eliminating the cavity mode to arrive at an effective spin model.  Relaxing both of these constraints provides access to rich physics encompassing coupled dynamics of atoms and light in the celebrated Dicke model~\cite{dicke1954coherence,dimer2007proposed,ritsch2013cold,kirton2019introduction,mivehvar2021cavity}:
\begin{equation}\label{eq:dicke}
H_\mathrm{D} = \omega_0 S_z + \omega_c a^\dagger a +  \mathcal{G}\left(a^\dagger + a\right)\left(S^+ + S^-\right),
\end{equation}
The key difference between the Dicke model and the Tavis-Cummings model of Eq.~\ref{eq:Htc} is the inclusion of counter-rotating terms signifying the correlated creation (or annihilation) of a photon and a spin excitation.  These terms are negligible for a two-level atom coupled to an optical cavity, where the mode frequency is much larger than the coupling $\mathcal{G}$, but can be engineered in Raman schemes~\cite{sorensen2002entangling,dimer2007proposed}.

A hallmark of the Dicke model is a quantum phase transition arising from the interaction between the collective spin component $S_x$ and the quadrature operator $X \propto a^\dagger + a$ of the cavity field.  Above a critical coupling strength $\mathcal{G}$, the phase of the light becomes correlated with the phase of the atoms, spontaneously breaking $\mathds{Z}_2$ symmetry.  In a common experimental implementation, the collective spin $\vec{S}$ is a pseudo-spin degree of freedom representing momentum modes of a cold atomic gas~\cite{black2003observation} or Bose-Einstein condensate~\cite{baumann2010dicke}.  A transverse drive field induces Raman processes, parameterized by the coupling $\mathcal{G}$, in which atoms change their momentum states by scattering photons into the cavity.  The symmetry-broken phase then manifests as an atomic density wave with one of two possible checkerboard orderings, which are correlated with two possible phases ($0$ or $\pi$) of the intracavity field.  Signatures of the Dicke phase transition have been observed both in the output light from the cavity~\cite{black2003observation} and in time-of-flight images of the atoms~\cite{baumann2010dicke}.

\begin{figure}[htb]
\includegraphics[width=\textwidth]{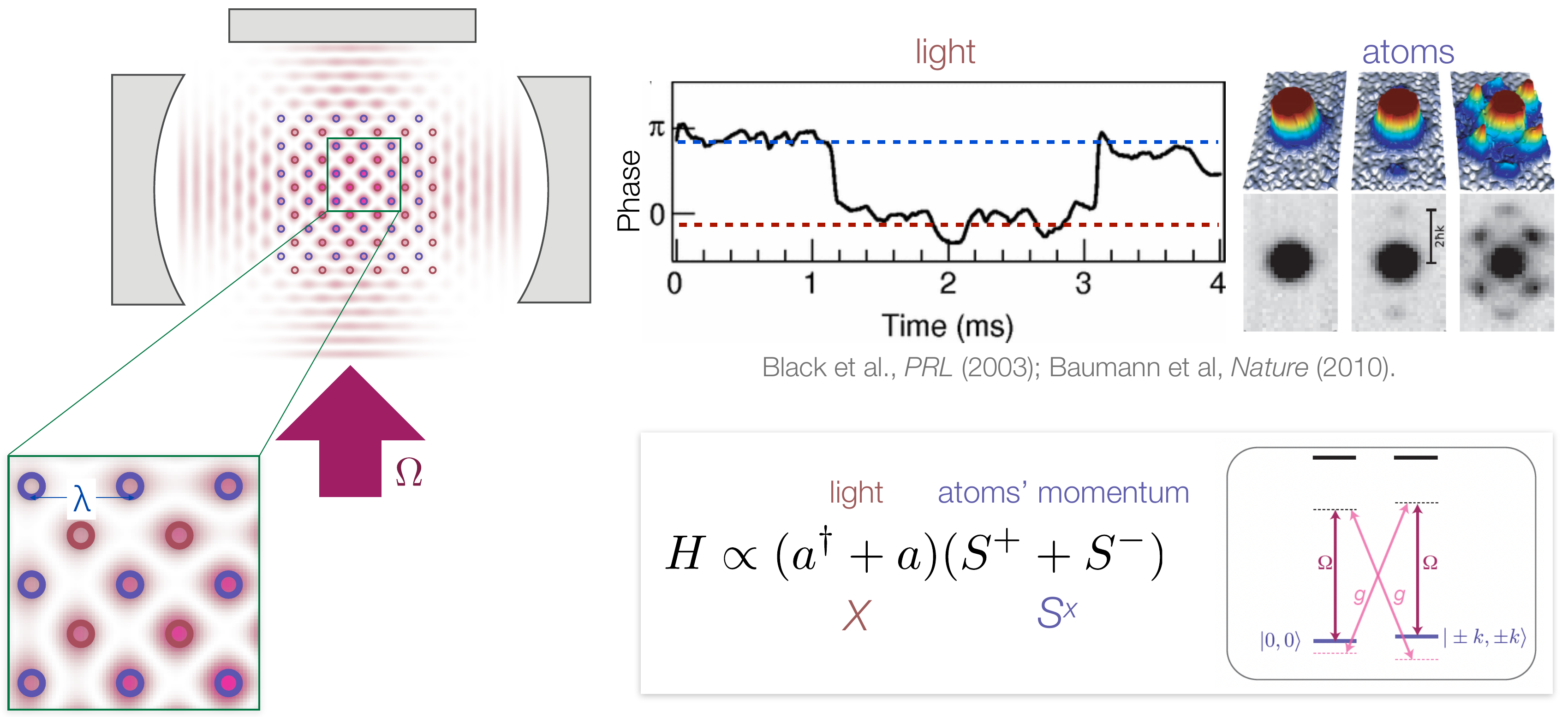}
\caption{\textbf{Self-organization in the Dicke model.}  Left: typical experimental setup featuring transversely driven cavity. Red and blue circles show the two possible checkerboard orderings of the self-organized atomic density wave.  Top right: signatures of the self-organized phase in the phase of the output light and in time-of-flight images.  Bottom right: Hamiltonian and atomic level scheme for pseudo-spin $S$ encoded in momentum states.}
\end{figure}

Following seminal observations of the Dicke phase transition in Refs.~\cite{black2003observation,baumann2010dicke}, a generalization to two crossed cavities enabled the realization of a $U(1)$-symmetric model and the observation of a self-organized density wave that spontaneously breaks this continuous symmetry, signifying a supersolid phase~\cite{leonard2017supersolid}.  Further extending this approach to a multimode cavity enabled the observation of phonons within the self-organized density wave~\cite{guo2021optical}.  Implementations of the Dicke model with Raman couplings between internal states have led to spin self-organization~\cite{kroeze2018spinor,landini2018formation}, while the canonical implementation with motional states has been extended to degenerate Fermi gases~\cite{zhang2021observation,helson2023density}.  Very recently, studies of the Dicke phase transition have been extended to a mesoscopic regime using control of ten to twenty individual atoms in optical tweezers~\cite{ho2025optomechanical}.

\subsection{Tailoring the Graph of Cavity-Mediated Interactions}
As a bus for mediating interactions between arbitrary atom pairs, a cavity is in principle a versatile tool for quantum simulation, notably enabling nonlocal connectivity.  Realizing this full potential requires transforming the native all-to-all couplings into a wider range of engineered graphs of cavity-mediated interactions.  Two complementary approaches to tailoring the coupling graph are Floquet engineering using time-dependent control fields; and leveraging multiple cavity modes.  To motivate these approaches, we will first examine the capabilities and limitations of a single-mode cavity, with only static control, for venturing beyond the effectively single-particle physics of a collective spin.

\subsubsection{Spatially Varying Coupling in Single- and Multimode Cavities}\label{sec:multimode}

We so far focused on the case where all atoms are uniformly coupled to a single cavity mode, leading to the collective spin models discussed in Sec.~\ref{sec:collective_spin}.  More generally, however, both the atom-cavity coupling $g_j$ and the Rabi frequency $\Omega_j$ of the drive field in a Raman scheme may depend on the atomic positions.  The spin-spin couplings are then of the form $J_{ij}\propto \mathcal{G}_i^* \mathcal{G}_j$, where $\mathcal{G}_j \propto \Omega_j^* g_j$.  Equivalently, the cavity couples to a nonuniformly weighted collective spin $\boldsymbol{\mathcal{F}} \propto \sum_i \mathcal{G}_i \vec{s}_i$, mediating interactions of the form
\begin{equation}
H_\mathrm{inh} \propto \sum_{\alpha,i,j} J_\alpha \mathcal{G}_i^* \mathcal{G}_j s^\alpha_i s^\alpha_j = \sum_\alpha J_\alpha \mathcal{F}_\alpha^2,
\end{equation}
where $\alpha \in \{x,y,z\}$ denotes a spin component.  The properties of such separable spin models have been analyzed theoretically in Refs.~\cite{bentsen2019integrable,marino2019cavity} and are discussed further in Sec.~\ref{sec:random_spin_models}.  However, a limitation for explorations of many-body physics is that the inhomogeneous couplings $\mathcal{G}_j$ offer only limited control of the coupling matrix, which remains restricted to a rank of unity.

A paradigm for accessing matrices of higher rank is to generalize to interactions $J_{ij}$ that are linear combinations of separable couplings:
\begin{equation}\label{eq:multimode}
J_{ij} \propto \sum_{\mu=1}^M \mathcal{G}_{\mu,i}^*\mathcal{G}_{\mu,j}.
\end{equation}
One way to achieve this employs a set of $M$ near-degenerate cavity modes indexed $\mu$, each with a different dependence $\mathcal{G}_{\mu,i}\propto g_{\mu,i}$ of the atom-cavity coupling on atomic position.  Letting these modes simultaneously mediate interactions allows for realizing more complex graphs than possible with static coupling to a single cavity mode.  The multimode approach was employed to tune the range of interactions mediated by a confocal cavity in a seminal experiment by Vaidya et al.~\cite{vaidya2018tunable}.  More recently, following pioneering proposals in Refs.~\cite{strack2011dicke,gopalakrishnan2011frustration}, a multimode cavity has been employed to realize a driven-dissipative spin glass~\cite{kroeze2025directly} (Sec.~\ref{sec:spin_glass}).

An analogous approach to engineering coupling matrices of arbitrary rank exploits trotterization or Floquet engineering~\cite{hung2016quantum,periwal2021programmable}, sequentially coupling a single cavity mode to $M$ different collective spin modes indexed $\mu$, e.g., $\mathcal{F}^+_\mu = \sum_i c_{\mu,i}s^+_i$.  The resulting time-averaged Hamiltonian has a coupling graph $J_{ij}\propto \sum_\mu c^*_{\mu,i}c_{\mu,i}$, equivalent to Eq.~\ref{eq:multimode}.  The following section (Sec.~\ref{sec:programmable}) elaborates on this approach, focusing first on an intuitive physical picture for realizing arbitrary translation-invariant interaction graphs by Floquet engineering.

\subsubsection{Programmable Nonlocal Interactions via Floquet Engineering}\label{sec:programmable}

A versatile approach to programming the graph of cavity-mediated interactions is provided by Floquet engineering with a modulated drive field~\cite{hung2016quantum,bentsen2019treelike,periwal2021programmable}.  In its simplest incarnation, this method allows for specifying the dependence of spin-exchange couplings on distance $r_{ij}$ between spins indexed $(i,j)$ to obtain an effective Hamiltonian
\begin{equation}\label{eq:Hprog}
\Heff = \frac{1}{2}\sum_{ij} \Jprog(r_{ij}) S^+_i S^-_j + \mathrm{h.c.}
\end{equation}
Here, $\vec{S}_j$ may denote either either the spin of a single atom or a collective spin formed by an atomic ensemble.

\begin{figure}[htb]
\includegraphics[width=\textwidth]{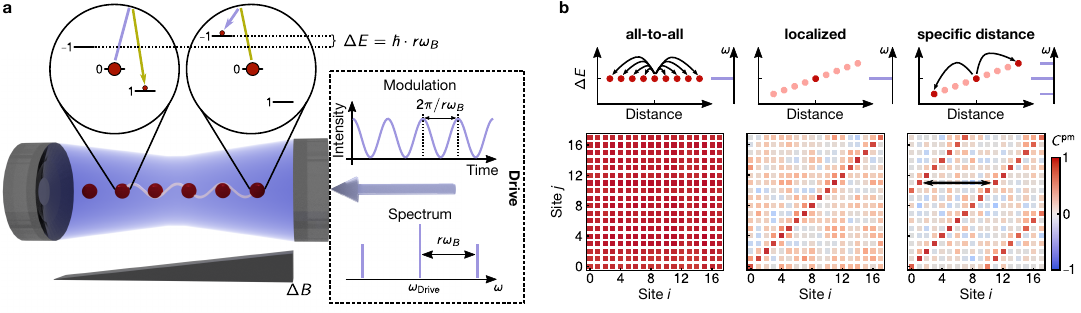}
\caption{\textit{Reproduced from Ref.~\cite{periwal2021programmable}}. (a) Scheme for programming the distance-dependence of cavity-mediated interactions via Floquet engineering with a modulated drive field.  (b) Measured spin correlations in an array of ensembles of spin-1 atoms.  Left: in the absence of the gradient, correlations extend across the entire array.  Center: turning on a magnetic field localizes the interactions to within individual sites, suppressing off-diagonal correlations.  Right: modulating the drive field at frequency $r\omega_B$ (with $r=10$) introduces interactions and corresponding correlations at the specified distance $r$.}\label{fig:programming_interactions}
\end{figure}

To engineer programmable XY models from native all-to-all spin-exchange couplings (Eq.~\ref{eq:Hxxz} with $J_z = 0$), a starting point is to introduce an inhomogeneous field $h_j = x_j\omega_B$ that depends linearly on the position $x_j$ of the $j^\mathrm{th}$ spin, described by $H_\mathrm{inh} = \sum_j h_j S^z_j$.  The field gradient suppresses cavity-mediated spin-exchange processes by imposing an energy penalty $r\omega_B$ for spin exchange between sites separated by a distance $r$.  Yet the interactions at distance $r$ can be controllably reintroduced by modulating the native global coupling $J_{xy}(t)$ --- controlled by the intensity of a drive field --- at frequency $r\omega_B$~(Fig.~\ref{fig:programming_interactions}).  More generally, a multitone modulation allows for turning on interactions at multiple distances, with the intensity and phase of each tone governing the magnitude and phase of the corresponding spin-exchange coupling.

An alternative picture for understanding the same method of programming interactions is in momentum space.  The magnetic field gradient provides a means of sequentially coupling the cavity to different spin-wave modes.  In particular, the magnetic field gradient induces Bloch oscillations for magnons, such that spin waves of momentum $k$, created by the operator $\mathcal{F}^+_k \propto \sum_j e^{i k j} S^+_j$, evolve in the Heisenberg picture as $\mathcal{F}^+_k(t) = \mathcal{F}^+_{k-\omega_Bt}(0)$.  In other words, each mode of a given momentum $k$ at $t=0$ transforms into the symmetric mode that couples to the cavity at a later time $t_k = k/\omega_B$.  In momentum space, the Hamiltonian thus takes the form $\Heff \propto \sum_k\mathcal{E}(k)\mathcal{F}^+_k \mathcal{F}^-_{-k}$, where the magnon dispersion relation $\mathcal{E}(k)$ at each momentum $k$ is directly set by the drive intensity at the corresponding time $t_k$~\cite{periwal2021programmable}.  

\begin{figure}[htb]
\includegraphics[width=\textwidth]{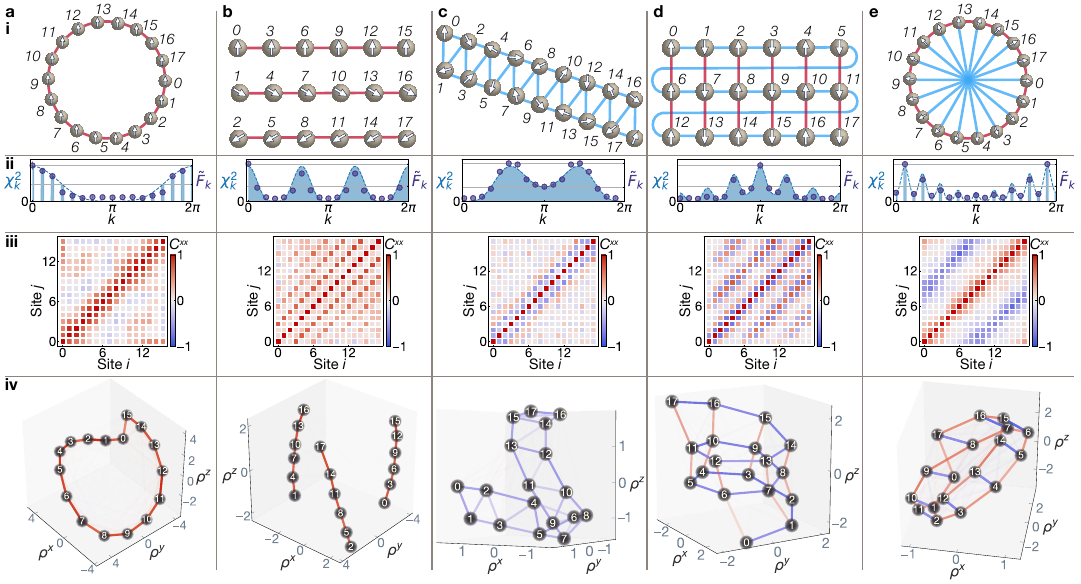}
\caption{\textit{Reproduced from Ref.~\cite{periwal2021programmable}}.  Gallery of interaction graphs programmed via Floquet engineering, and black-box reconstruction of the coupling graphs from measured spin correlations.}\label{fig:programmable_graphs}
\end{figure}

Programmable cavity-mediated interactions were demonstrated in Ref.~\cite{periwal2021programmable} in the setting of an array of atomic ensembles.  Here, despite the one-dimensional physical geometry of the array, we engineered a variety of interaction graphs best visualized by an embedding in two or three spatial dimensions, as shown in Fig.~\ref{fig:programmable_graphs}.  The experiment operated with spin-1 atoms, which we intialized in the $m=0$ Zeeman state.  A cavity drive field was then turned on to perform a quench under the programmable spin-exchange interactions in Eq.~\ref{eq:Hprog}.  In the presence of a quadratic Zeeman shift $q > 0$, and for ferromagnetic coupling $J_{xy}(t)$ in the native Hamiltonian, the system is unstable to forming correlated atom pairs in $m = \pm 1$ (see Sec.~\ref{sec:spin_mixing}).  Since populating these states increases the quadratic Zeeman energy, conservation of energy requires that the interaction energy $H_{xy}$ decrease.  The resulting spin correlations resemble those expected for a low-temperature state of a classical XY model on the programmed interaction graph~\cite{periwal2021programmable}.

For example, Fig.~\ref{fig:programmable_graphs}(e) shows spin correlations measured with the interaction graph $\Jprog(r_{ij})$ being a M\"obius ladder with sign-changing interactions.  The graph is formed by couplings at distances $r\in\{1,9,18\}$ in a chain of $N=18$ sites, realized by modulating the drive field at the corresponding frequencies $r\omega_B$.  The phase of the modulation (0 or $\pi$) governs the sign of the coupling, which was chosen to be ferromagnetic for the rails ($r=1$) and antiferromagnetic for the rungs ($r=9$).  

To verify the programmed coupling graph, we interpreted the spin correlations $C^{xx}_{ij} \propto \avg{F^x_iF^x_j}-\avg{F^x_i}\avg{F^x_j}$ as a measure of distance $d_{ij}$ in an effective geometry.  Adopting the physically motivated ansatz that correlations decay as a Gaussian function $\abs{C^{xx}} \propto e^{-d_{ij}^2}$ of the effective distance $d_{ij}$, we found the best-fit embedding of sites in three dimensions to match the distance matrix $d_{ij}$, shown in Fig.~\ref{fig:programmable_graphs}(iv).  Each reconstructed graph shows bonds between \textit{all} $N(N-1)/2$ pairs of sites, with color and opacity indicating of the sign and strength of coupling inferred from measured correlations.  The results visually corroborate the successful realization of the target interaction graphs.

The Floquet approach to programming interactions generalizes beyond the translation-invariant graphs considered in Refs.~\cite{hung2016quantum,periwal2021programmable}. Since the cavity is (approximately) agnostic to the physical position of the atoms, the coupling graph is solely determined by the local fields $h_j$ combined with the spectrum of the drive field.  Thus, greater flexibility is readily attainable by optically imprinting the local fields~\cite{cooper2024graph}, which need not depend linearly on site index $j$.  In a complementary picture, recall that purpose of the linear field gradient was to transform between spin-wave modes, which form a basis in which any translation-invariant Hamiltonian is diagonal.  By extension, \textit{any} graph can be engineered by diagonalizing the coupling matrix to find a basis of collective modes to couple to the cavity, as we shall further discuss in Sec.~\ref{sec:graph_states}.

\section{Applications in Quantum Metrology}\label{sec:metrology}

A primary driver of advances in cavity QED with atomic spin ensembles has been the goal of enhancing the precision of atomic clocks and sensors by entanglement.  Compared with alternative approaches to generating entanglement, cavity-based techniques offer several advantages for metrology.  Firstly, both the absolute precision and the enhancement attainable from entanglement grow with increasing atom number, motivating methods of scalably generating entanglement among many atoms.  Additionally, while generating correlations requires interactions, either among the atoms or between the atoms and a probe, the ability the turn these interactions off after entanglement generation is crucial for minimizing systematic effects in a precision measurement.  Thus, the collective, optically switchable interactions enabled by coupling atoms to a cavity offer an ideal toolbox for quantum metrology.

\subsection{Squeezed Atomic Clocks and Sensors}
Cavity-based spin squeezing has been applied to enhance the precision of atomic clocks and atom interferometers in proof-of-principle experiments~\cite{leroux2010implementation,hosten2016measurement,pedrozo2020entanglement}, which have employed both measurement- and interaction-based approaches to generating entanglement.  Building on these demonstrations, an ongoing effort seeks to incorporate entanglement into state-of-the-art devices to obtain a true quantum advantage over the best performance possible with classical techniques.

\begin{figure}[htb]
\includegraphics[width=\textwidth]{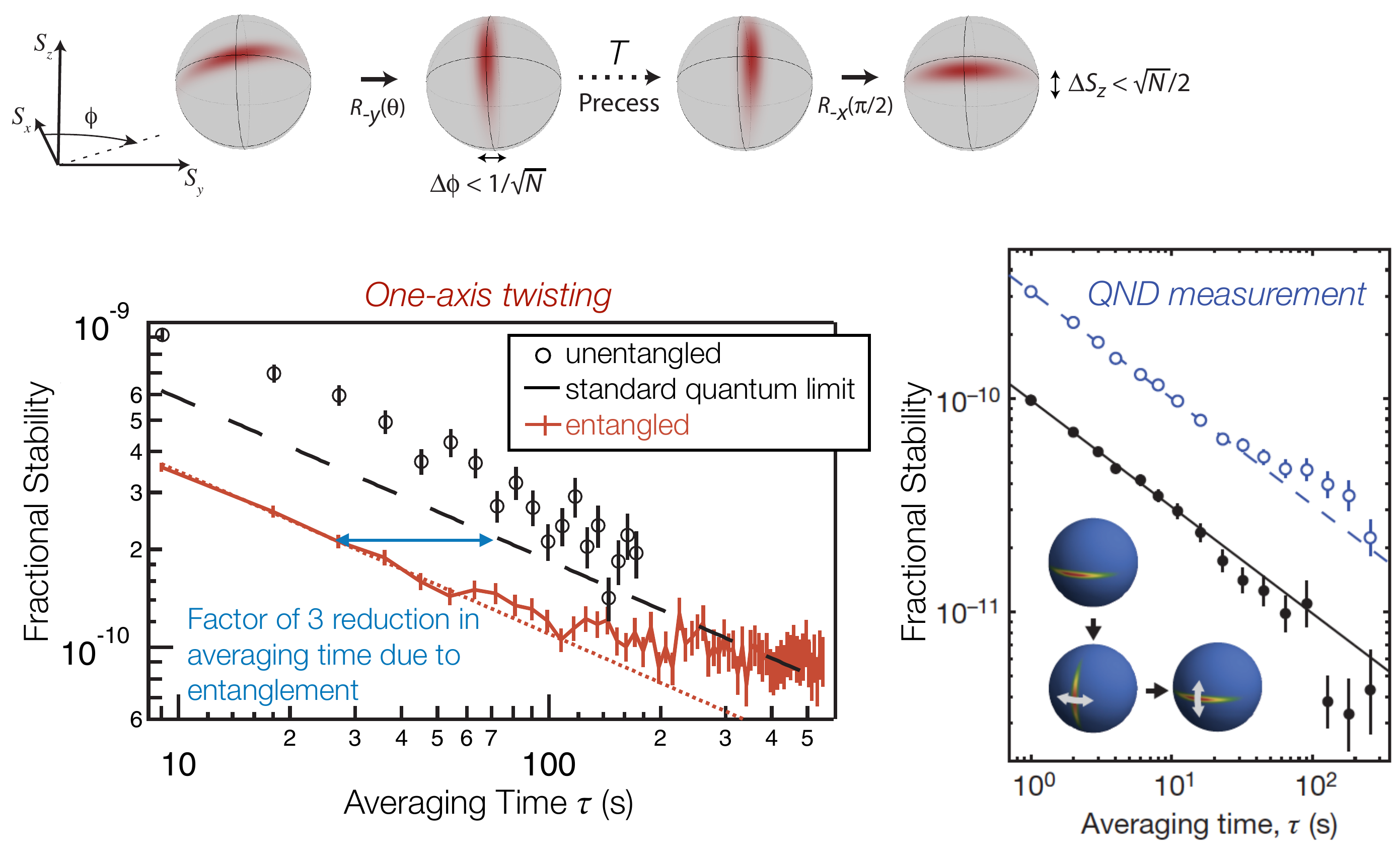}
\caption{\textbf{Squeezed clocks.}  Top: Schematic of Ramsey spectroscopy with a squeezed spin state.  Bottom: Fractional frequency stability $\sigma$ of squeezed microwave clocks from Refs.~\cite{leroux2010orientation,hosten2016measurement}.}\label{fig:squeezed_clocks}
\end{figure}

To understand the requirements for enhancing the performance of state-of-the precision measurements, consider the case of an atomic clock limited only by fundamental quantum noise.  A canonical protocol for operating the clock is Ramsey spectroscopy, as illustrated in Fig.~\ref{fig:squeezed_clocks}.  After repeatedly performing Ramsey spectroscopy over a total averaging time $\tau$, the fractional frequency stability (Allan deviation)
\begin{equation}
\sigma(\tau) = \frac{\xi}{\sqrt{N}\omega T}\sqrt{\frac{T_\mathrm{cycle}}{\tau}}
\end{equation}
depends on the atom number $N$, the transition frequency $\omega$, the duration $T$ of a single interrogation and cycle time $T_\mathrm{cycle}\ge T$ between interrogations, and the metrological squeezing parameter $\xi$ defined in Eq.~\ref{eq:wineland_parameter}.  Whereas the first demonstrations of clock stability beyond the standard quantum limit operated on a microwave transition in rubidium, with clock transition frequency $\omega=2\pi\times 6.8$~GHz~\cite{leroux2010orientation,hosten2016measurement}, recent experiments have boosted the stability by orders of magnitude by applying squeezing to optical clock transitions in ytterbium and strontium~\cite{pedrozo2020entanglement,robinson2024direct}.  Still, microwave-transition clocks remain of interest for applications requiring portability, motivating a recent demonstration of squeezed Ramsey spectroscopy in a microchip-based rubidium clock with a record second-scale interrogation time~\cite{huang2023observing}.

The interrogation time is a a crucial consideration for attaining a practical metrological gain from squeezing.  If this time is limited by atom loss or decoherence that degrades the squeezing, initializing the system in an entangled state provides at best a marginal benefit~\cite{huelga1997improvement}.  However, in the realistic scenario where a limit is set by the coherence time of the local oscillator (LO) --- e.g., the clock laser in an optical atomic clock --- squeezing provides an advantage by allowing the LO phase error to be determined more precisely before incurring a phase-wrapping ambiguity~\cite{andre2004stability,pezze2020heisenberg}.  A caveat is that verifying this benefit requires performing a clock comparison, since the LO no longer provides a reliable reference once its noise exceeds the quantum noise $\Delta\phi = \xi/\sqrt{N}$.  Thus, an important milestone was a recent comparison of two squeezed optical clocks within a single optical cavity, which allowed for quantifying the clock stability at the $10^{-17}$ level~\cite{robinson2024direct}.

Spin squeezing has also been applied in a proof-of-principle demonstration of an entanglement-enhanced matter-wave interferometer~\cite{greve2022entanglement}.  For this application, the atoms were placed in free-fall within an optical cavity, and the initial coherent spin state was prepared using a velocity-sensitive Raman transition that placed the atoms in a superposition $\ket{\downarrow}=\ket{f=1,+2\hbar k},\ket{\uparrow}=\ket{f=2,-2\hbar k}$ of internal (hyperfine) and momentum states.  Both measurement-based squeezing and one-axis twisting were successfully applied to realize interferometric sensitivity beyond the standard quantum limit.

\subsection{Enhanced Sensing via Time Reversal}\label{sec:time_reversal}

\begin{figure}[htb]
\includegraphics[width=\textwidth]{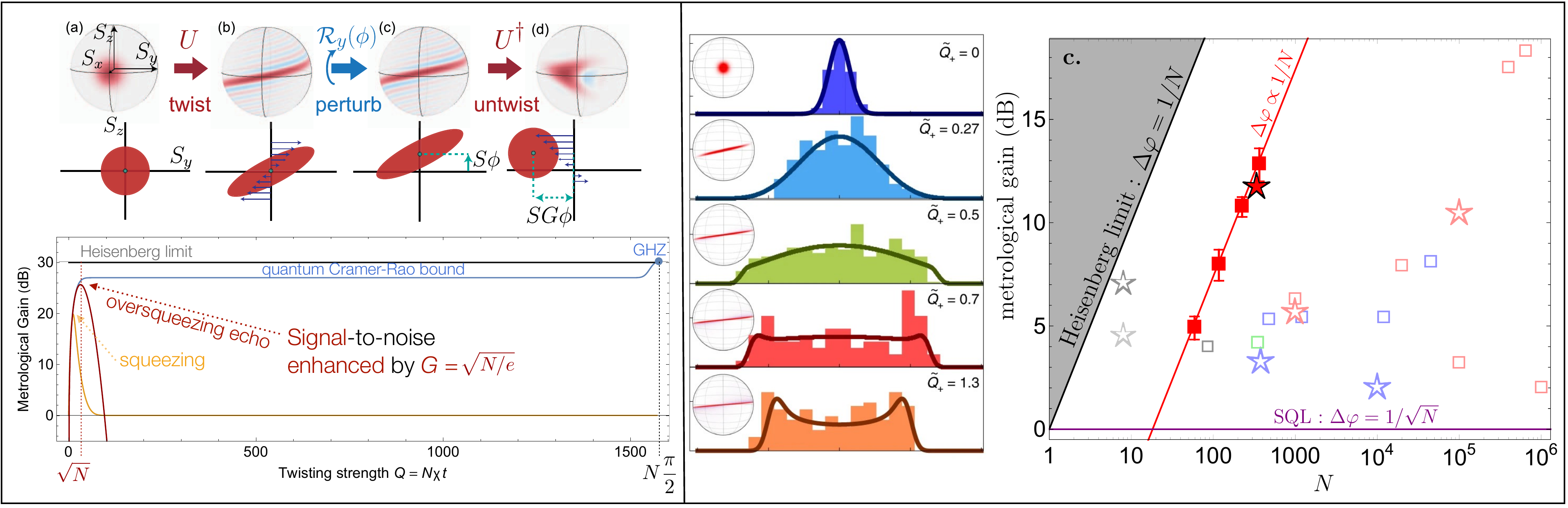}
\caption{\textbf{Time reversal for enhanced metrological gain from one-axis twisting.} Figure adapted from theoretical proposal in Ref.~\cite{davis2016approaching} (left) and from experimental demonstration in Ref.~\cite{colombo2022time}.}\label{fig:time_reversal}
\end{figure}

In practice, the metrological gain attained from cavity-mediated entanglement may be limited by factors other than finite cooperativity.  One common limitation is detection noise: the stronger the squeezing, the more challenging it becomes to resolve the quantum fluctuations.  In particular, gaining the full benefit of the entanglement requires that the technical noise of the readout be negligible compared with the squeezed quantum noise.  Additionally, for interaction-based squeezing via one-axis twisting, the state becomes non-Gaussian at late times (Fig.~\ref{fig:time_reversal}).  Although the intrinsic sensitivity to perturbations continues to grow, as quantified by the quantum Fisher information~\cite{giovannetti2011advances,pezze2018quantum}, the standard protocol of Ramsey spectroscopy fails to extract the full metrological benefit of the oversqueezed, non-Gaussian state~\cite{davis2016approaching}.

To overcome limitations posed by both detection noise and non-Gaussianity, a powerful approach is to use interactions not only to generate an entangled state but also to facilitate its readout.  One broadly applicable protocol is an echo comprising evolution $U=e^{-iH_I t}$ under an interaction Hamiltonian $H_I$, application of the perturbation $\mathcal{R_\alpha}$ (e.g., a spin rotation about the $\alpha$-axis) that one wishes to sense, and a subsequent ``time-reversal'' step in which the sign of interactions is switched to evolve under $-H_I$.  In the net evolution $U^\dagger \mathcal{R}_\alpha U$, the perturbation $\mathcal{R}_\alpha$ manifests in a failure to return to the initial state, as illustrated in Fig.~\ref{fig:time_reversal} for the case of an initial coherent spin state subject to one-axis twisting ($H_I = \Htwist$)~\cite{davis2016approaching}.  The requisite time reversal is natural to implement using cavity-mediated interactions, since the sign of the spin-spin couplings is controlled by the sign of a laser detuning that can readily be switched.

The benefit of the echo is most intuitively understood in the case where the interactions ($U$) simply squeeze the quantum noise.  In this case, the reverse evolution $U^\dagger$ amplifies both the quantum noise and the signal, converting the small displacement of the squeezed state into a large displacement of the initial coherent state that is easier to detect.  This amplification has been demonstrated in experiments using cavity-mediated one-axis twisting, and variously dubbed ``quantum phase magnification''~\cite{hosten2016quantum} or ``signal amplification by time-reversed interaction (SATIN)''~\cite{colombo2022time}.  For early-time dynamics where the state remains Gaussian, time reversal is not strictly necessary for the signal amplification step and can equivalently be replaced by performing a spin rotation followed by continued forward evolution.  This approach was employed in Ref.~\cite{hosten2016quantum} to attain phase sensitivity 8~dB beyond the standard quantum limit, despite operating with detection noise 10 dB \textit{above} the projection noise level.

A more fundamental benefit of interaction-based readout for one-axis twisting emerges in the regime of oversqueezing at times $\chi t \sim \sqrt{N}$, where the state becomes non-Gaussian~\cite{davis2016approaching}.  This non-Gaussianity limits the squeezing to an optimum value $\xi^2 \propto N^{-2/3}$~\cite{kitagawa1993squeezed}, yet the state's intrinsic metrological potential continues to grow~\cite{davis2016approaching}.  Specifically, a generalized squeezing parameter $\xi^2_\QFI = N(\Delta\phi)^2 = N/\QFI_\alpha$ quantifying the sensitivity to rotations $\mathcal{R}_\alpha(\phi)=e^{i\phi S_\alpha}$ can be defined based on the quantum Fisher information (QFI) $\QFI_\alpha = 4(\Delta S_\alpha)^2$ for a pure state~\cite{giovannetti2011advances,pezze2018quantum}.  The entangled states generated by one-axis twisting permit a Heisenberg scaling $\xi^2_\QFI \propto 1/N$ for times $\chi t\sim \sqrt{N}$, and the echo sequence provides a straightforward protocol for extracting this scaling in practice, by producing a final state that is approximately a coherent state rotated from its initial orientation by an amount $G\phi$, with amplification factor $G\sim \sqrt{N}$.

The Heisenberg scaling of the one-axis twisting echo was observed in an experiment employing cavity-mediated interactions among up to $N=400$ atoms in Ref.~\cite{colombo2022time}, yielding a record 11.8(5) dB enhancement in spectroscopic sensitivity due to entanglement.  This success in harnessing the oversqueezed non-Gaussian state for metrological gain relied on several key conditions: the capacity for switching the sign of interactions via the sign of a laser detuning; operation in the regime of single-atom strong coupling $\eta > 1$; and the use of a cycling transition to attain a metrological gain scaling as $\xi^{-2} \propto N\eta$ (as discussed in Sec.~\ref{sec:qnd_measurement}).

The use of time reversal for metrology is closely related to probes of many-body chaos, including the Loschmidt echo~\cite{jalabert2001environment} and out-of-time-order correlation functions~\cite{larkin1969quasiclassical,maldacena2016bound,swingle2016measuring}, which quantify sensitivity to perturbations and the ``butterfly effect’’ in which a local perturbation spreads under the influence of interactions.  Inspired by this connection, time-reversal metrology has been extended to a system with exponential sensitivity to perturbations, namely, the Lipkin-Meshkov-Glick model comprising global Ising interactions in a transverse field~\cite{li2023improving}.  The observed metrological gain due to entanglement was compared with an out-of-time-order correlation function quantifying quantum information scrambling.

\subsection{Programmable Multimode Entanglement: Graph States}\label{sec:graph_states}

We have so far focused on quantum sensing applications in which the goal is to estimate just a single parameter, such as the accumulated phase in an atomic clock or, analogously, the strength of a global magnetic field.  However, a variety of applications --- including vector magnetometry~\cite{kaubruegger2023optimal}, spatially resolved sensing~\cite{muessel2014scalable}, optimal quantum clocks~\cite{pezze2020heisenberg} and distributed clock networks~\cite{malia2022distributed} — benefit from access to multimode entanglement.  This motivates the question: how can we control the spatial structure of entanglement among multiple atomic ensembles?

Recent work has shown that the combination of global cavity mediated interactions with local addressing suffices to flexibly engineer the structure of multimode squeezing in an array~\cite{cooper2024graph}.  For example, whereas the cavity most naturally generates global squeezing, the addition of local control alternatively allows for localizing squeezing to within individual ensembles or generating two-mode squeezing that maximizes the Einstein-Podolsky-Rosen (EPR) correlations \textit{between} two ensembles.  More broadly, the same toolbox provides flexible control over a wide class of \textit{continuous-variable graph states}, named for a graph that specifies the structure of entanglement between nodes formed by the atomic spin ensembles.

The setting for the demonstration of programmable entanglement in Ref.~\cite{cooper2024graph} is an optical cavity featuring an array of ensembles of rubidium-87 atoms, with $N\approx 10^3$ atoms per ensemble.  The ensembles are loaded from an array of optical microtraps into an intracavity lattice of 1560-nm wavelength, designed to provide approximately uniform coupling to the 780-nm cavity mode that mediates interactions.  Specifically, Ref.~\cite{cooper2024graph} employed interactions among spin-1 atoms giving rise to spin nematic squeezing (Sec.~\ref{sec:spin_mixing}), but the scheme for programming entanglement is agnostic to the particular method of cavity-mediated squeezing employed.  Indeed, we may think of each ensemble abstractly as (approximately) an oscillator with conjugate variables $(x_j, p_j)$ (Fig.~\ref{fig:spin_mixing}), such that an atomic ensemble initialized in a coherent spin state corresponds to the oscillator being initialized in its vacuum state.  Generically, our goal might be to squeeze the fluctuations of each individual ensemble or to generate correlations between the ensembles.

\begin{figure}[htb]
\includegraphics[width=\textwidth]{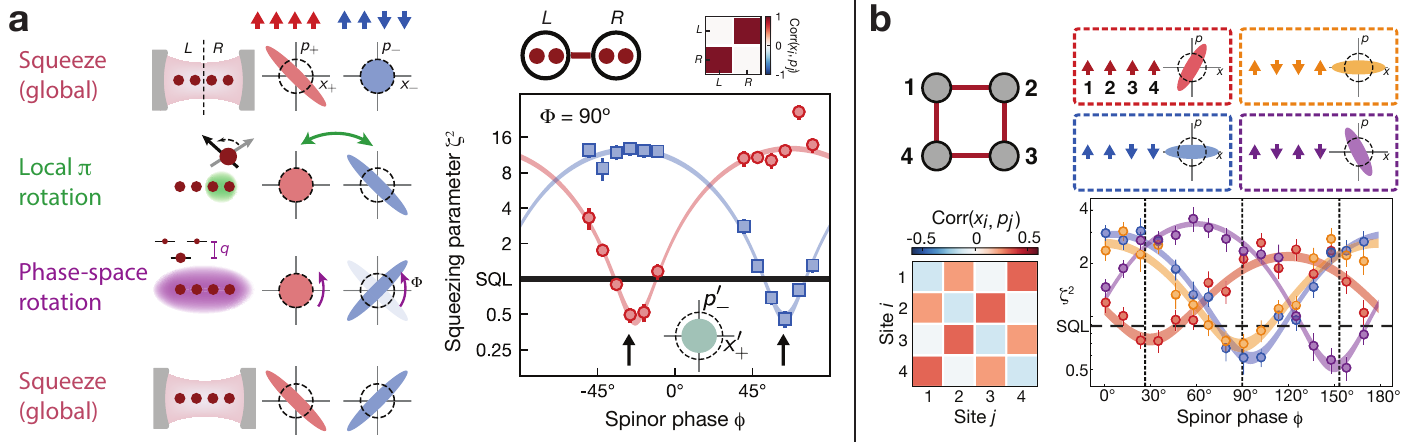}
\caption{\textbf{Programming the spatial structure of entanglement} by combining global cavity-mediated interactions with local addressing.  \textit{Figure adapted from Ref.~\cite{cooper2024graph}}. (a) Preparation of EPR entangled state by squeezing the symmetric (red) and antisymmetric (right) modes with relative phase $\Phi=\pi/2$ between the squeezed quadratures for the two modes.  (b) Preparation of a four-mode square graph state by squeezing four collective modes.}\label{fig:epr_and_square_graph}
\end{figure}

A versatile approach to engineering spatial structure of the squeezing is to operate on a basis of \textit{collective modes}.  For an array of $M$ sites, the system can be fully described in a basis of $M$ collective modes $(X_\mu, P_\mu)$ indexed $\mu$, which are related to the local degrees of freedom by a linear transformation $V$, e.g., $X_\mu = V_{\mu j}x_j$ (where we sum over the repeated index $j$).  The collective mode that couples to the cavity is the ``center-of-mass'' (symmetric) mode, with quadratures $X_+ = \sum_j x_j/\sqrt{M}$ and $P_+ = \sum_j p_j/\sqrt{M}$.  However, the squeezing can be transferred to another collective mode via local operations.  For example, in the minimal case of $M=2$ sites, the squeezing can be mapped from the symmetric mode ($X_+ \propto X_1 + X_2$) to the antisymmetric mode ($X_- \propto X_1 - X_2$) by applying a local $\pi$ rotation to one of the two sites [Fig.~\ref{fig:epr_and_square_graph}(a)].

Whereas squeezing the symmetric mode is useful for sensing a global field, squeezing the antisymmetric mode instead maximizes sensitivity to a field gradient.  Further, if we seek to know \textit{both} the average field and the gradient (equivalently, to measure the field locally on each site), we should squeeze both modes.  Indeed, squeezing the symmetric and antisymmetric modes in the same quadrature is equivalent to producing two independent squeezed states, generating correlations within each subsystem.  By contrast, squeezing these two collective modes in orthogonal quadratures is equivalent to producing a two-mode squeezed (EPR entangled) state that maximizes correlations \textit{between} subsystems.  The EPR entanglement constitutes a resource for the counterintuitive task of simultaneously sensing perturbations of noncommuting observables to arbitrary precision, effectively ``evading quantum mechanics''~\cite{tsang2012evading} to surpass the local Heisenberg uncertainty bound.

Einstein-Podolsky-Rosen (EPR) entanglement, as well as the opposite limit of local squeezing, was generated in Ref.~\cite{cooper2024graph} by combining global squeezing with local control, as shown in Fig.~\ref{fig:epr_and_square_graph}(a).  The protocol consists of driving a cavity to induce global squeezing; applying a $\pi$ spin rotation to one of two sites to transform the antisymmetric mode into the symmetric one that couples to the cavity; and subsequently applying a second global squeezing operation.  In the simplest case, the local $\pi$ rotation acts to echo away the interactions between the two systems, such that the second squeezing operation removes intersystem entanglement while reinforcing intrasystem entanglement.  However, if the local $\pi$ pulse is accompanied by a global $\pi/2$ phase space rotation that switches the squeezed and antisqueezed quadratures, the net effect is to echo away the interactions \textit{within} each subsystem while reinforcing the interactions \textit{between} subsystems.

More generally, operating in the basis of collective modes enables efficient generation of a broad class of entangled states known as \textit{graph states}.  In systems of qubits, such states are of interest as a universal resource for measurement-based computation and are prepared by applying a controlled-$Z$ gate on each edge of a graph or, equivalently, evolving under an Ising Hamiltonian $\sum_{i,j} A_{ij} \sigma^z_i \sigma^z_j$, where $A_{ij}$ is the \textit{adjacency matrix} of the graph; the resulting state features correlations between $\sigma_z$ on a given site and $\sigma_x$ on neighboring sites.  An analogous class of states in multimode bosonic systems are \textit{continuous-variable} graph states prepared by evolving an initially unentangled system under interactions of the form $H \propto \sum_{ij}A_{ij} x_i x_j$, yielding correlations between $x$ on a given site and $p$ on the neighboring sites.

A general prescription for preparing a target graph state via global cavity-mediated interactions is obtained by diagonalizing the adjacency matrix $A_{ij}$.  One thereby obtains a set of eigenvectors specifying collective modes to squeeze, while each corresponding eigenvalue $\lambda_\mu$ indicates which quadrature should be squeezed.  For example, an EPR entangled state of two modes has adjacency matrix
\begin{equation}
A = \begin{pmatrix}
0 & 1 \\ 1 & 0
\end{pmatrix},
\end{equation}
with eigenvectors $\vec{v}_\pm = (1, \pm 1)/\sqrt{2}$ representing the symmetric and antisymmetric modes; and eigenvalues $\lambda_\pm = \pm 1$ specifying that the collective modes should be squeezed in quadratures $x\cos\phi_\pm  + p\sin\phi_\pm$ with $\phi_\pm = \arctan(\lambda_\pm)=\pm\pi/4$, i.e., in orthogonal quadratures.  In Ref.~\cite{cooper2024graph}, where we applied this prescription to prepare a two-mode EPR entangled state, we additionally scaled it to preparing a four-mode square graph state [Fig.~\ref{fig:epr_and_square_graph}(b)] via a sequence of four collective squeezing operations interspersed by local operations that transform between modes.

The technique for engineering multimode entanglement in the collective mode basis is fundamentally scalable.  Arbitrary $M$-mode graphs are accessible via $M$ collective squeezing operations, each of which occurs at a collectively enhanced rate to yield a fixed total interaction time.  Further, the attainable squeezing is fundamentally limited only by the collective cooperativity per mode~\cite{cooper2024graph}.  As the degree of squeezing in Fig.~\ref{fig:epr_and_square_graph} is predominantly limited by technical noise sources including imperfect fluorescence detection, in ongoing work we are leveraging interaction-based readout to maximize the metrological gain attainable from programmable multimode entanglement~\cite{cooper2024programmable}.

The ability to flexibly program the spatial structure of entanglement opens the door to exploring advanced protocols in quantum metrology.  As one indication of the potential power of multimode Gaussian states, recent theoretical work has shown that the sensitivity of EPR entangled states to perturbations in non-commuting observables opens the door to attaining exponential quantum speedups in learning certain spatiotemporally correlated noise processes~\cite{oh2024entanglement}.  Additional opportunities include leveraging multimode control for optimal vector magnetometry~\cite{kaubruegger2023optimal}, where cavity-mediated interactions can enable both entanglement generation and optimal decoding via interaction-based readout; and for entanglement-enhanced compressed sensing of spatially patterned fields~\cite{baamara2023quantum}.

\section{Applications in Quantum Simulation}

The nonlocal connectivity provided by a cavity is, in principle, a powerful resource for quantum simulation.  Realizing this potential requires venturing beyond the collective spin systems most naturally realized in a single-mode cavity, which are confined to a small Hilbert space amenable to efficient classical simulation.  We have already introduced key tools for escaping this narrow realm, including modifying the native global couplings of a single-mode cavity by Floquet engineering in an inhomogeneous field (Sec.~\ref{sec:programmable}), employing local optical addressing (Sec.~\ref{sec:graph_states}), or incorporating multiple cavity modes (Sec.~\ref{sec:multimode}).  We now turn to a few examples of ongoing and prospective applications of this toolbox for quantum simulation of models inspired by condensed-matter physics and quantum gravity.

\subsection{Dynamics of Nonlocal Spin Systems}\label{sec:random_spin_models}

One broad source of inspiration for quantum simulations with cavity-mediated interactions is that nonlocal couplings are intimately connected with fundamental bounds in quantum dynamics.  This connection is epitomized by the paradigmatic Sachdev-Ye (SY) model of spins with random all-to-all Heisenberg couplings~\cite{sachdev1993gapless,sachdev2023quantum}
\begin{equation}
\HSY = \frac{1}{\sqrt{N}}\sum_{i,j}J_{ij}\vec{s}_i\cdot\vec{s}_j 
\end{equation}
and its generalization to fermions with random multibody interactions, the Sachdev-Ye-Kitaev (SYK) model~\cite{kitaev2015simple,maldacena2016remarks}.  The latter is a maximally chaotic fast scrambler~\cite{sekino2008fast}, holographically dual to a black hole under the anti de Sitter / conformal field theory (AdS/CFT) correspondence~\cite{maldacena2016remarks}.

The original random Heisenberg magnet $H_\mathrm{SY}$ is best known as a putative quantum spin glass at low temperature.  While realizing such a quantum spin glass would be of interest in its own right~\cite{strack2011dicke,gopalakrishnan2011frustration}, there is furthermore evidence that the Sachdev-Ye model $H_\mathrm{SY}$ exhibits SYK-like features including maximal chaos at temperatures above the spin-glass transition~\cite{sachdev2023quantum}. Numerical studies also indicate that the low-temperature glassiness can be averted in variants of the spin model with multibody couplings~\cite{swingle2024bosonic}, and recent cavity-QED experiments have even realized three- and four-body interactions in collective spin systems~\cite{luo2024realization}.

The Sachdev-Ye model $H_\mathrm{SY}$ is superficially similar to the collective Heisenberg spin model introduced in Sec.~\ref{sec:tunable_heisenberg}, yet a key difference lies in the randomness of the couplings.  In particular, realizing a full-rank coupling matrix $J_{ij}$ for a system of $N$ spins requires either employing an extensive number $O(N)$ of cavity modes to mediate interactions; or sequentially coupling $O(N)$ collective atomic modes to a single-mode cavity to Floquet-engineer the couplings.  While neither of these options appears prohibitive for modest system sizes $N\lesssim 100$, the technical and fundamental challenges are significant enough to warrant asking whether simpler models can access similar physics.  Both theory~\cite{marino2019cavity,bentsen2019integrable,bentsen2019treelike,kim2020low} and experiments~\cite{davis2020engineering,sauerwein2023engineering,periwal2021programmable,kroeze2025directly} have begun to investigate this question.

A minimal ingredient for accessing many-body spin dynamics in a single mode cavity is to incorporate spatial inhomogenity that breaks permutation symmetry.  In this spirit, a natural class of models to investigate takes the form
\begin{equation}\label{eq:Hxxzh}
\Hxxzh = \left[J_{xy}(\F_+ \F_- + \F_- \F_+) + J_z \F_z^2 \right] 
\end{equation}
where $\boldsymbol{\mathcal{F}} = \sum_j c_j \vec{f}_j$ represents a collective spin that is weighted to account for nonuniform couplings $w_j$ to the cavity.  A key feature of this model is its integrability for isotropic Heisenberg couplings ($J_z = J_{xy}$) --- where it belongs to a well-studied class of so-called Richardson-Gaudin models~\cite{sklyanin1989separation} --- and in the pure XY ($J_z=0$) case~\cite{bentsen2019integrable}.  Beyond this regime, the model is predicted to exhibit a rich phase diagram as a function of anisotropy $J_z/J_{xy}$ and of the spin length $f$, including regions that are chaotic in the classical limit ($f\rightarrow \infty$) but nearly integrable for quantum spins (e.g, $f=1/2$)~\cite{bentsen2019integrable}.  Cavity experiments in principle offer unique capabilities for tuning between the quantum and classical limits, by operating with random weights $w_i$ for individual spin-$f$ atoms or by placing groups of $n$ atoms into supersites forming larger spins of length $nf$.

Even with uniform atom-cavity couplings, a nonuniform external field provides an alternative means of accessing many-body spin dynamics~\cite{davis2020protecting,sauerwein2023engineering,young2024observing}.  Notably, a simplified model of Bardeen-Cooper-Schrieffer (BCS) superconductivity is realized by global spin-exchange interactions in an inhomogeneous field~\cite{anderson1958random,young2024observing},
\begin{equation}
H_{XY,h} = \frac{J_{xy}}{2}\sum_{ij}\left(s^+_i s^-_j + \mathrm{h.c.}\right) + \sum_j h_j s^z_j,
\end{equation}
the same native Hamiltonian that served the starting point for Floquet-engineered interactions in Sec.~\ref{sec:programmable}.  The relationship of $H_{XY,h}$ to BCS superconductivity is given by the Anderson pseudospin mapping~\cite{anderson1958random}, where each spin (with $s=1/2$) represents a momentum mode in which the raising operator $S^+_i$ creates a Cooper pair and the local fields $h_j$ specify the dispersion relation.  Under this mapping, a recent experiment simulated dynamical phases of a BCS superconductor by probing mean-field dynamics of an atomic ensemble in a cavity~\cite{young2024observing}.  One feature of this phase diagram is that, above a critical mean field interaction strength, the interactions protect the coherence of the collective spin, as also observed in Ref.~\cite{davis2020protecting}.  The same model $H_{XY,h}$ was also studied in an experiment by Sauerwein \textit{et al.}, which examined the influence of disordered fields $h_i$ on the localization of eigenstates~\cite{sauerwein2023engineering}.

Future work may both directly simulate the SYK model~\cite{baumgartner2024quantum}, building on recent experiments with fermions in cavities~\cite{helson2023density,sauerwein2023engineering}, and explore the scrambling dynamics of a wider class of nonlocally interacting systems that pose a challenge to numerical simulations.  Toward this end, demonstrated tools of Floquet engineering (Sec.~\ref{sec:programmable}) and dynamical optical addressing (Sec.~\ref{sec:graph_states}) offer a potentially powerful toolbox.  One proposed route to fast scrambling is to combine cavity-mediated global interactions with time-dependent random local fields~\cite{belyansky2020minimal}. Section~\ref{sec:holographic} further introduces an alternative class of putative fast scrambler, featuring a sparse, deterministic pattern of nonlocal interactions~\cite{bentsen2019treelike} that we engineered experimentally in Ref.~\cite{periwal2021programmable}.  While existing experiments that leverage collectively enhanced atom-light coupling are already well positioned to study scrambling dynamics in regimes amenable to semiclassical simulation, accessing strongly interacting quantum dynamics remains an outstanding challenge discussed further in Sec.~\ref{sec:stronger_coupling}.

\subsection{Spin Glasses and Optimization}\label{sec:spin_glass}

Spin glasses --- disordered, frustrated magnets whose signatures include absence of conventional magnetic order, non-ergodicity, and aging --- are of long-standing interest within and beyond physics~\cite{stein2013spin}.  In a classic series of essays in \textit{Physics Today}, P.W. Anderson summarized their significance~\cite{anderson1988spin}: \textit{``The pursuit of the spin glass mystery led, inter alia and aside from all the good solid-state physics that resulted, to new algorithms for computer optimization, a new statistical mechanics, a new view of protein structure, a new view of evolution and new ideas in neuroscience.''}  By today, the success of artificial intelligence could be added to this list, punctuated by Nobel Prizes awarded to J. Hopfield for his model of associative memory as an Ising spin glass; and to G. Parisi for uncovering an entirely new form of symmetry breaking characterizing the spin glass phase.  Still, remaining mysteries --- particularly surrounding the nature of quantum spin glasses --- motivate the investigation of spin glasses in quantum simulators~\cite{strack2011dicke,gopalakrishnan2011frustration,marsh2024entanglement,kroeze2025directly}.

Paradigmatic theoretical models of spin glasses, including the SY model introduced in Sec.~\ref{sec:random_spin_models}, feature nonlocal interactions~\cite{strack2011dicke,gopalakrishnan2011frustration,kroeze2025directly}, with a mix of ferromagnetic and antiferromagnetic couplings contributing to frustration.  In condensed-matter systems, this nonuniform sign arises when conduction electrons mediate interactions between magnetic impurities, leading to the Ruderman, Kittel, Kasuya, Yoshida (RKKY) interaction~\cite{ruderman1954indirect}.  Seminal proposals in Refs.~\cite{gopalakrishnan2011frustration,strack2011dicke} pointed out that cavities can mediate a similar form of interaction, thanks to the spatial oscillation in the mode function.

For concreteness, consider the Raman scheme for generating cavity-mediated spin-exchange couplings in Fig.~\ref{fig:spin_exchange}.  For a uniform drive field (incident transversely) of Rabi frequency $\Omega$ and a standing-wave cavity mode with dependence $g(\vec{r}) \approx g_0 \cos(kr)$ on position $\vec{r}$, the couplings mediated by this single cavity mode are of the form
$J_{ij}^\mathrm{M=1} \propto \cos(\vec{r}_i)\cos(\vec{r}_j)$.
While seemingly sign-changing, these interactions are not yet frustrated, since local changes of basis suffice to make all of the couplings ferromagnetic.  Yet incorporating multiple degenerate cavity modes, as described in Sec.~\ref{sec:multimode}, breaks the separability of the couplings to produce frustration.

\begin{figure}[htb]
\includegraphics[width=\textwidth]{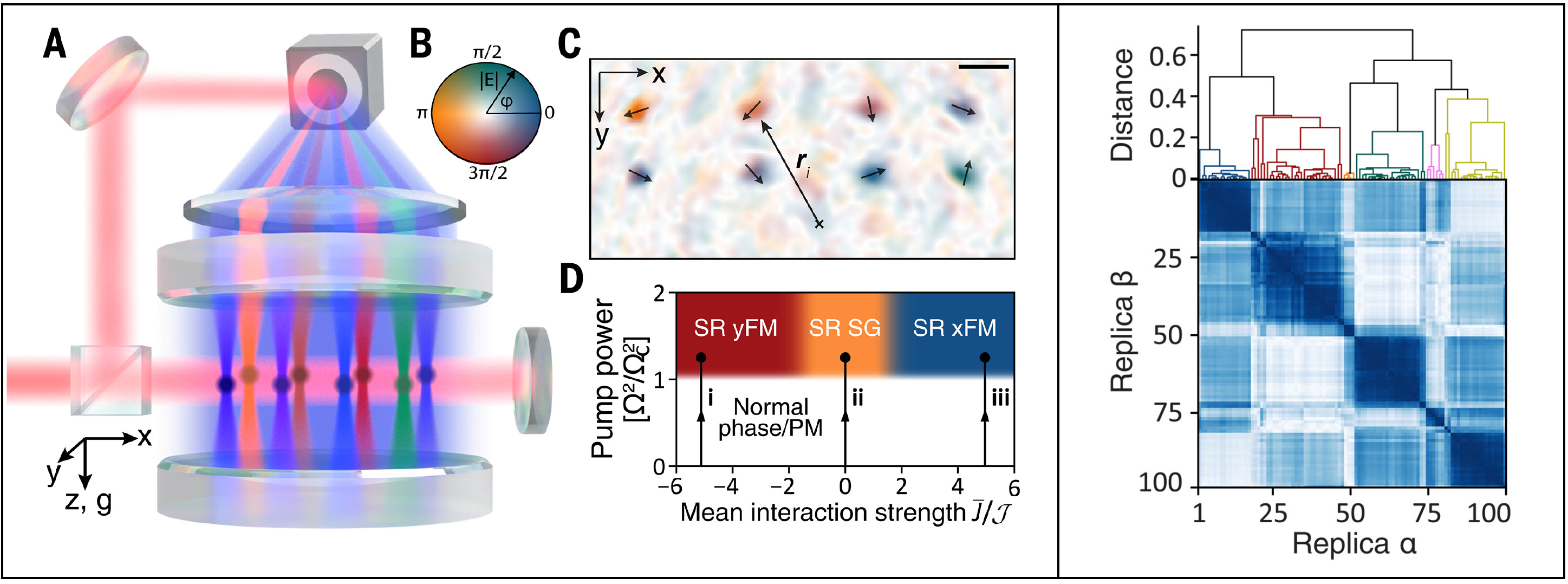}\caption{\textbf{Vector spin glass in a multimode cavity.} \textit{Figure adapted from Ref.~\cite{kroeze2018spinor}}. Left: experimental platform featuring an array of eight atom clouds in a transversely pumped confocal optical cavity.  Right: ultrametricity evidenced by a hierarchical structure of correlations between replicas.}\label{fig:spin_glass}
\end{figure}

A variant of this approach enabled the observation of a vector spin glass in a multimode cavity~\cite{kroeze2025directly}.  In this pioneering experimental demonstration, the spins were encoded in the phases of an array of ultracold atom clouds that were optically trapped within a confocal cavity (Fig.~\ref{fig:spin_glass}).  The cavity was driven transversely to realize a multimode generalization of the Dicke model, leading to local self-organized density waves.  Viewing the phase of the density wave within each site as a vector spin, the cavity mediated non-local, sign-changing interactions between these spins.

A highlight of the experiment by Kroeze \textit{et al.} in Ref.~\cite{kroeze2025directly} was the direct observation of replica symmetry breaking, a form of order that appears only in correlations between multiple copies of identically prepared systems.  Analyzing the correlations between replicas revealed a hierarchical structure signifying an emergent ultrametricity in the space of spin configurations --- i.e., different low-energy configurations may be viewed as situated on a ``family tree'' indicating a scale-invariant structure of basins of attraction in the rugged energy landscape.

The same rugged energy landscape underlies the computational complexity of a wide range of optimization problems that map to energy minimization in nonlocal spin models.  As a key feature of the cavity-QED realization of a spin glass in Ref.~\cite{kroeze2025directly} is its driven-dissipative nature, theoretical work has begun to explore the role of the open-system dynamics, as well as the role of entanglement, in finding low-energy states.  Deep in the strong coupling regime, cavity-mediated interactions may also enable explorations of optimization algorithms that rely crucially on maintaining coherence.  For example, mapping the NP hard problem of number partitioning to a system of spins with weighted couplings to a single cavity mode theoretically enables hardware-efficient algorithms based on Grover search~\cite{anikeeva2021number} or on quantum annealing~\cite{ye2023universal}.  

\subsection{Holographic Duality: Emergent Geometry from Entanglement}\label{sec:holographic}

Can experimental advances in controlling and probing entanglement with atoms in cavities offer a new window into theories of quantum gravity?  Lest this idea sound far-fetched, I will contextualize it with a quotation from the father of the atom, Democritus: \textit{``Sweet is by convention and bitter is by convention, hot by convention, cold by convention, color by convention, in truth there are but atoms and the void''}~\cite{graham2010texts}.  In modern language, Democritus is remarking on the \textit{emergence} of taste, color, and texture from more fundamental building blocks of atoms.  The end of his remark touches on the void, which still holds mystery today for philosophers and physicists alike.  Could spacetime itself also be an emergent phenomenon, and what might be its microscopic building blocks?

An ever expanding body of theoretical work is exploring the notion that entanglement forms the underlying fabric of spacetime.  Quantifying this notion is the Ryu-Takayanagi formula~\cite{ryu2006holographic}, which relates entanglement entropies in a $d$-dimensional quantum system to the areas of minimal surfaces in a $(d+1)$-dimensional geometry known as the \textit{holographic bulk}.  By distilling certain strongly correlated quantum systems into simpler descriptions in terms of curved space and gravity, theories of holographic duality --- notably the AdS/CFT correspondence~\cite{maldacena1999large} --- may offer a new handle for understanding and calculating properties of the quantum many-body system.  However, a challenge in pursuing this vision is the paucity of theoretically tractable models of holographic duality.  Those models that do have known holographic duals -- such as the SYK model discussed in Sec.~\ref{sec:random_spin_models} — involve exotic, nonlocal interactions, posing a challenge to the common paradigm of benchmarking a quantum simulator in a well understood setting before venturing into the unknown.

Experiments with atoms in cavities offer a promising avenue for exploring holographic duality in the lab by offering some of the features of holographic models: a starting point is the native nonlocal connectivity, and recent experiments have further introduced multibody interactions~\cite{luo2024realization} and fermionic degrees of freedom~\cite{helson2023density}.  In the spirit of starting with models that look ``simple’’ on the gravitational side, one prospect is to directly simulate the SYK model — dual to a (0+1)-dimensional black hole.  Broader opportunities range from understanding which models beyond SYK exhibit black-hole-like fast scrambling; to expanding beyond $d=0$ dimensions in the quantum system to explore the duality between entanglement and geometry.

\begin{figure}[tbh]
    \centering
    \includegraphics[width=\textwidth]{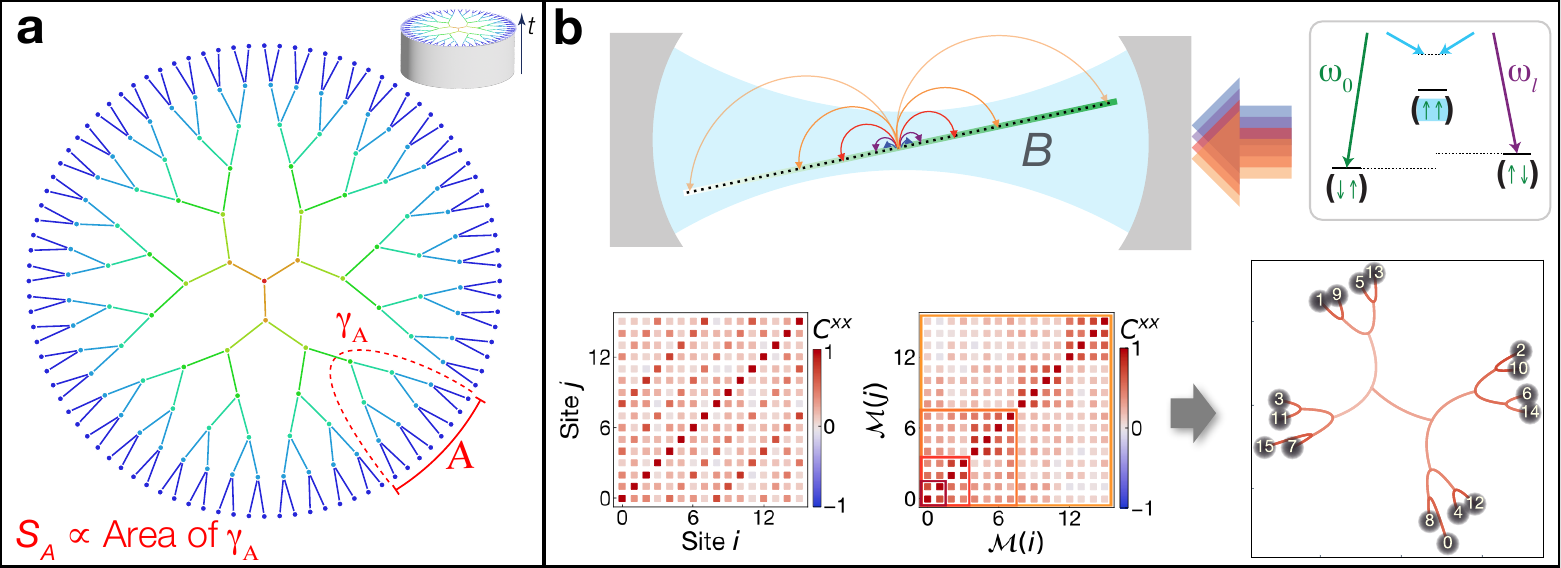}
    \caption{\textbf{(a) Relationship between entanglement and geometry.}  The Ryu-Takayanagi (RT) formula relates the entanglement entropy $S_A$ of a region $A$ of a quantum system (dark blue sites on boundary) to the area of the minimal surface $\gamma_A$ (dashed red curve) subtending the boundary region $A$. \textbf{(b) Emergent holographic bulk geometry in a cold-atom quantum quantum simulator.} \textit{Adapted from Refs.~\cite{bentsen2019treelike,periwal2021programmable}}. Top: scheme for programming interactions among atomic ensembles in a magnetic field gradient via the frequency spectrum of a drive field, as described in Sec.~\ref{sec:programmable}.  Bottom: measured spin correlations $C^{xx}_{ij}$ and reconstruction of emergent treelike bulk geometry~\cite{periwal2021programmable}.  The emergent treelike bulk geometry corresponds to a hierarchical structure of the correlation matrix, which becomes evident when sites are rearranged according to a map $\mathcal{M}(i)$ that reverses the order of bits in a binary representation of the site index $i$.
}
    \label{fig:cavity_treelike}
\end{figure}

These goals motivated us to investigate a toy model related to a version of holographic duality known as $p$-adic AdS/CFT, in which the holographic bulk has an explicit representation as a tree graph forming a discretization of hyperbolic (AdS) space~\cite{gubser2017p,heydeman2016tensor,bhattacharyya2018tensor}.  To realize a quantum system that effectively resides on the boundary of such a tree, we engineered a sparse pattern of nonlocal couplings
\begin{equation}
J(r) = \begin{cases}
r^{s}&r=2^\ell\,\,(\text{for integer } \ell = 0, 1, 2, \dots)\\0&\text{otherwise}
\end{cases}
\end{equation}
at power-of-two distances $r$ in an array of atomic spin ensembles~\cite{bentsen2019treelike,periwal2021programmable}.  Here, the power-law exponent $s$ tunes between two radically different geometries: for $s<0$, where nearest-neighbor interactions dominate, the system is simply a chain with trivial bulk geometry.  For $s>0$, where \textit{furthest}-neighbor interactions dominate, the interactions give rise to a hierarchical structure of correlations [Fig.~\ref{fig:cavity_treelike}(b)] suggestive of a treelike bulk geometry.  Although the interior vertices of the tree are unphysical in the quantum system, we observed the emergence of the treelike bulk geometry by applying a coarse-graining procedure to the measured spin correlations: drawing bonds between the most strongly correlated sites, grouping these into supersites, drawing bonds between strongly correlated supersites, and repeating the process until the system was fully connected revealed the tree in Fig.~\ref{fig:cavity_treelike}(b).

Our reconstruction of the treelike bulk geometry used only on measurements of spin correlations in a single basis, and thus so far did not directly witness or characterize entanglement in the system.  However, we have since characterized the spatial structure of entanglement in continuous-variable graph states in the same platform~\cite{cooper2024graph}, as described in Sec.~\ref{sec:graph_states}.  Future work may thus aim to directly reconstruct emergent bulk geometries from experimental measures of entanglement~\cite{cao2020building}.  Here, the multimode Gaussian states arising from the early-time dynamics of cavity-mediated interactions among collective spins offer a useful test case, as entanglement entropy can be efficiently reconstructed from measured correlations.

The tree graph defines a \textit{non-Archmidean geometry}, in which the distance $\abs{i-j}_2$ between a pair of sites $i,j$ (the $2$-adic norm) is measured by counting how many levels one must go up the tree to connect them.  Importantly, sites that are close to each other on the tree (i.e., strongly correlated for $s>0$) are far apart in the physical (Archimedean) geometry.  At the transition point $s=0$ between the two geometries, all notion of locality breaks down, yielding strong connectivity that is conducive to efficiently delocalizing information.  Indeed, numerical calculations in a semiclassical regime show a logarithmic scrambling time $t_* \propto \log N$~\cite{bentsen2019treelike}.

The putative fast scrambler at $s=0$ is markedly different from the paradigmatic SYK model, featuring sparse (rather than dense) connectivity and deterministic (rather than random) couplings.  Experimentally exploring the dynamics of this and other models accessible via cavity-mediated nonlocal connectivity~\cite{belyansky2020minimal}, with the aid of demonstrated techniques for probing information scrambling~\cite{swingle2016measuring,lewis2018unifying,li2023improving}, thus offers prospects for elucidating necessary and sufficient conditions for realizing holographic duals of black holes.

\subsection{Topological Order, Error Correction, and Novel Spin Liquids}

Topological phases of matter are of fundamental and practical interest for their capacity to store quantum information robustly in nonlocal degrees of freedom.  This robustness is necessarily in tension with the requirement of manipulating and accessing such nonlocally encoded information when needed.  An inspiring theoretical proposal by Jiang \textit{et al.} envisioned leveraging a cavity to controllably map nonlocal degrees of freedom onto ancilla qubits as a means of manipulating and probing logical qubits in a topological quantum memory~\cite{jiang2008anyonic}.  For example, cavity-mediated Ising interactions between an ancilla qubit and a set $\mathcal{S}$ of data qubits in principle allow for mapping the Pauli string $\prod_{j\in{\mathcal{S}}} \sigma^z_j$ onto the ancilla.  Such operators also are of interest for implementing quantum low-density parity check (LDPC) codes, which reduce the overhead for quantum error correction at the cost of requiring measurements of nonlocal stabilizer operators~\cite{chandra2024non}.

One major challenge that must be addressed to realize these visions is boosting the coherence of cavity-mediated interactions.  A promising avenue is to harness a high-cooperativity millimeter-wave cavity, building on demonstrated successes in preparing and probing nonclassical field states in such a cavity via strong dispersive coupling~\cite{deleglise2008reconstruction} (Sec.~\ref{sec:qnd_photons}).  Such a cavity offers the prospect of high-fidelity photon-mediated gates and, thanks to the long wavelength, provides ample space for trapping large two-dimensional Rydberg atom arrays with uniform coupling to the mode~\cite{zhang2025optically}.

While one prospect is to use a cavity to probe and manipulate topological matter generated by demonstrated techniques of local Rydberg interactions~\cite{semeghini2021probing,bluvstein2022quantum}, cavity QED also offers a unique toolbox for accessing topologically ordered states.  In a landmark experiment, minimal instances of photonic Laughlin states were prepared in a twisted optical cavity~\cite{clark2020observation}.  The twist produced a strong effective magnetic field for the photons, which were furthermore induced to interact by hybridization with Rydberg excitations in an atomic gas.  This work highlighted the capacity of cavity-QED systems to explore strongly correlated matter in a driven-dissipative setting.

In future experiments, photon-mediated interactions offer prospects for realizing engineered and emergent topological phases in long-lived atomic spin degrees of freedom.  One canonical route to engineering topologically ordered states, well established for ground states of the toric code, involves preparation of a graph state and measurement of the subset of the qubits~\cite{bluvstein2022quantum}.  The cavity-based method of preparing graph states presented in Ref.~\cite{cooper2024graph} and Sec.~\ref{sec:graph_states} may offer generalizations to a wider range of graphs than those naturally realized by local gates and atom movement.

A complementary paradigm is to seek emergent strongly correlated states by engineering flat bands for magnons via frustrated spin-exchange couplings.  Towards this end, the programmable interactions described in Sec.~\ref{sec:graph_states} and Ref.~\cite{periwal2021programmable} offer a high degree of flexibility for engineering the band structure, by permitting arbitrarily long-range couplings as well as arbitrary phases of coupling matrix elements.  A further prospect is to combine local Rydberg interactions with cavity-mediated nonlocal interactions, with the latter inducing frustration in settings where it would ordinarily be absent.  In particular, a cavity-mediated global antiferromagnetic interaction can energetically constrain a system to the highly degenerate space of singlet pairings.  Thus, even on an otherwise unfrustrated square lattice, Ref.~\cite{chiocchetta2021cavity} has predicted that a cavity-mediated global Heisenberg coupling stabilizes a spin-liquid phase.

\subsection{Challenges and Opportunities}\label{sec:stronger_coupling}
We have seen that atom-light interactions in optical cavities offer unique opportunities for quantum simulation of spin systems, including flexible control over nonlocal interactions and a natural setting for exploring driven-dissipative phenomena.  A caveat is that experiments to date have largely been confined to a realm that is efficiently numerically simulable, including studies of mean-field dynamics and quantum fluctuations amenable to semiclassical descriptions.

The underlying challenge is the limit set by the cooperativity $\eta$ on the interaction-to-decay ratio $J/\gamma \sim \sqrt{n\eta}$ for photon-mediated interactions between either single spin-$s$ atoms or $n$-atom ensembles representing larger spins $S=ns$.  Thus, at the cooperativities $\eta \lesssim 10^2$ realized to date in optical cavities, we are faced with a tradeoff between operating deep in the quantum regime ($S\sim 1$) at limited coherence or operating with larger spins $S\gg 1$.  Despite successes in characterizing entanglement between collective spins~\cite{cooper2024graph} and achieving sufficiently coherent dynamics to access non-Gaussian states~\cite{barontini2015deterministic,colombo2022time}, higher cooperativity will be required to access coherent dynamics in a strongly interacting quantum regime.

A promising route to addressing this challenge is to marry techniques of cavity QED and Rydberg physics.  One direction is to directly exploit the high cooperativity of millimeter wavelengths that couple directly to transitions between Rydberg states~\cite{zhang2025optically}, building on the demonstrated cooperativity $\eta>10^8$ achieved with circular Rydberg atoms~\cite{kuhr2007ultrahigh,deleglise2008reconstruction}.  A complementary paradigm compatible with optical cavities is to leverage stroboscopic Rydberg dressing~\cite{hines2023spin,weckesser2025realization,cao2025autoionization} or Rydberg-blockaded superatoms to combine collectively enhanced cooperativity with strong nonlinearity, up to the limit of spin-1/2 degrees of freedom~\cite{vaneecloo2022intracavity,ningyuan2016observation,clark2020observation}.

Additionally, recent advances in single-atom control and detection within optical cavities open prospects for entering a regime of quantum advantage via measurement.  Fluorescence imaging offers a close analogy with photon-number resolving detection in photonic platforms, which provides a resource for attaining quantum advantage even when otherwise restricted to Gaussian operations~\cite{madsen2022quantum}.  Cavity-aided quantum non-demolition measurements---providing opportunities for detecting nonlocal observables and for real-time feedback---offer prospects for further enriching the quantum simulation toolbox.

\begin{acknowledgement}
I gratefully acknowledge discussions with current and past members of my research group, including Emily Davis, Greg Bentsen, Eric Cooper, Avikar Periwal, and Philipp Kunkel.  I have also benefited from numerous discussions over the years with Vladan Vuletic, Mark Kasevich, Benjamin Lev, Jon Simon, Dan Stamper-Kurn, James Thompson, and additional colleagues in the cavity QED community whose work is cited throughout these notes.  Work from my own group reviewed in these notes was supported by the NSF, DOE, AFOSR, ARO, ONR, and Gordon and Betty Moore Foundation.
\end{acknowledgement}

%
%
\bibliographystyle{spphys.bst}
\bibliography{references}

\end{document}